%% file: ms.tex
\documentclass[twocolumn,tighten]{aastex61}

\usepackage{natbib,amssymb,url}
\include{mydefs}

\newcommand{\revise}{}

\providecommand{\sorthelp}[1]{}

\shortauthors{Friesen et al.}
\accepted{November 7, 2018}

\begin{document}

\title{ALMA detections of the youngest protostars in Ophiuchus}

\author{R. K. Friesen}
\affiliation{National Radio Astronomy Observatory, 520 Edgemont Rd., Charlottesville, VA, 22903, USA}
\author{A. Pon}
\affiliation{Department of Physics and Astronomy, The University of Western Ontario, 1151 Richmond Street, London, N6A 3K7, Canada}
\author{T. L. Bourke}
\affiliation{SKA Organisation, Jodrell Bank Observatory, Lower Withington, Macclesfield SK11 9DL, UK}
\author{P. Caselli}
\affiliation{Max-Planck-Institut f\"ur extraterrestrische Physik, Giessenbachstrasse 1, 85748 Garching, Germany}
\author{J. Di Francesco}
\affiliation{Herzberg Astronomy and Astrophysics, National Research Council of Canada, 5071 West Saanich Road, Victoria, BC, V9E 2E7, Canada}
\affiliation{Department of Physics and Astronomy, University of Victoria, 3800 Finnerty Road, Victoria, BC, Canada V8P 5C2}
\author{J. K. J{\o}rgensen} 
\affiliation{Centre for Star and Planet Formation, Niels Bohr Institute \& Natural History Museum of Denmark, University of Copenhagen, \O ster Voldgade 5-7, 1350, Copenhagen K., Denmark}
\author{J. E. Pineda}
\affiliation{Max-Planck-Institut f\"ur extraterrestrische Physik, Giessenbachstrasse 1, 85748 Garching, Germany}

\correspondingauthor{R. K. Friesen}
\email{rfriesen@nrao.edu}

\begin{abstract}

We present Atacama Large Millimeter/submillimeter Array (ALMA) observations of 1.1\ mm dust continuum and CO 2-1 emission toward six dense cores within the Ophiuchus molecular cloud. We detect compact, sub-arcsecond continuum structures toward \revise{three} targets, \revise{two} of which (Oph A N6 and SM1) are located in the Ophiuchus A ridge. \revise{Two targets, SM1 and GSS 30, contain two compact sources within the ALMA primary beam. We argue that} several of the compact structures are small ($R \lesssim 80$\ au) accretion disks around young protostars, \revise{due to} their resolved, elongated structures\revise{, coincident radio and x-ray detections, or} bipolar outflow detections. \revise{While CO line wings extend to $\pm 10-20$~\kms\ for the more evolved sources GSS 30 IRS3 and IRS1, CO emission toward other sources, where detected, only extends a few \kms\ from the cloud \vlsr.} The dust spectral index toward the compact objects suggests that the disks are either optically thick at 1.1~mm, or that significant grain growth has already occurred. \revise{We identify, for the first time,} a single compact continuum source ($R \sim 100$~au) \revise{toward N6} embedded within a larger continuum structure. \revise{SM1N is extended in the continuum but is highly centrally concentrated, with a density profile that follows a $r^{-1.3}$ power law within 200~au, and additional structure suggested by the uv-data. Both N6 and SM1N} show no clear bipolar outflows with velocities greater than a few km s$^{-1}$ from the cloud velocity. These sources are candidates to be the youngest protostars or first hydrostatic cores in the Ophiuchus molecular cloud. 

\end{abstract}


\section{Introduction}
\label{sec:intro}

Stars form within dense cores of molecular gas. Observations of the continuum emission from dust reveal that the radial density profiles of many starless cores can be well modeled with compact, flat centers that steepen to a power-law decrease at some radius. Cores that appear highly concentrated are more likely to contain embedded protostars \citep{walawender_2005}. In analytic equilibrium models, such as Bonnor-Ebert spheres \citep{bonnor_1956,ebert_1955}, this behaviour is expected as there is a maximum concentration value for critically self-gravitating objects beyond which a core is unstable to collapse \citep{johnstone00}. As the central density of a collapsing core increases beyond $\sim 10^{11}$\ \cc, the continuum emission becomes optically thick and the core begins to warm. The heated central molecular core achieves hydrostatic equilibrium, called the first hydrostatic core \citep[FHSC; ][]{larson69}, until H$_2$ is dissociated at $\sim 2000$\ K, with the subsequent collapse forming the second hydrostatic core, or protostar. Concurrently, infalling material may form a pseudo disk of a few hundred au in size around a young protostar, due to the influence of magnetic fields \citep[e.g.,][]{galli_shu1,galli_shu2} and/or rotation \citep[e.g.,][]{commercon_2012,bate_2014}.

\revise{Theoretical models of FHSCs predict a range of values for their masses, luminosities, sizes, and lifetimes given different assumptions regarding the magnetic field strength, rotation, and accretion rates \citep[see][for an overview of model predictions]{dunham_2014}. Without rotation, the predicted maximum masses and sizes are $M \sim 0.04 - 0.05$~\msun\ and $R \sim 5$~au, respectively \citep{boss_1995,masunaga98,saigo06,omukai_2007,tomida10}. Expected lifetimes range from $\sim 0.5 - 50$~kyr , or only a fraction of the Class 0 timescale \citep[$\sim 150$~kyr;][]{dunham_2014}. Rotation allows a broader range of FHSC masses, $M \sim 0.01 - 0.1$~\msun, in flattened, potentially disk-like cores with radii up to $\sim 10 - 20$~au \citep{bate98,saigo08,machida_2010,commercon_2012}. With predicted luminosities ranging from $\sim 10^{-4} - 10^{-1}$~L$_\odot$, FHSCs likely have little observable emission below $\sim 20 - 50$~\micron\ as they are deeply embedded in their natal cores \citep{boss_1995}. \citet{commercon_2012a} show, however, that at late times in their evolution, FHSCs may be detectable through sensitive 24~\micron\ observations.}  

FHSCs are also expected to drive slow, wide-angled outflows, with speeds of only a few \kms, in contrast to older protostars that drive more energetic outflows, although \citet{price_2012} show that faster ($\sim 7$\ \kms), better collimated outflows are possible given sufficiently high ionization. 
\revise{Over the FHSC lifetime range listed above, a 2~\kms\ outflow would reach a distance of only $\sim 200$~au to $\sim 0.1$~pc. }
Consequently, high spatial resolution of both continuum and kinematic tracers are needed to probe these prestellar stages of core evolution, and early stages of star formation. 

\revise{Recently, several sets of potentially very young protostellar objects have been characterized using sensitive infrared through millimeter and submillimeter imaging. These include a number of FHSC candidates \citep{enoch_2010,chen_2010,pineda11,pezzuto12,chen_2012,murillo_2013b}, protostars with internal luminosities $L_{int} \lesssim 0.1$~L$_\odot$ \citep[Very Low Luminosity Objects or VeLLOs;][]{difrancesco_2007}, and extremely red sources in the Orion molecular cloud consistent with young protostars embedded within high density cores \citep[PACS Bright Red Sources, or PBRS;][]{stutz_2013}. FHSC candidates are usually very faint or undetected at wavelengths shorter than 70~\micron. Where undetected in the mid- to far-infared, the presence of an embedded source is inferred via observations of compact submillimeter or millimeter continuum emission and/or low-velocity outflows. The set of VeLLOs may include some very young protostellar sources or FHSC candidates, but likely also include more evolved protostars with low accretion rates, and protobrown dwarfs \citep{dunham_2014}. The PBRS sample are more luminous than VeLLOs, but a subset are very young Class 0 objects, embedded in material with steep density profiles based on millimeter continuum analysis, and likely in a rapid accretion phase \citep{tobin_2015}.}

The transition from prestellar core to FHSC or protostar, while clearly a vital step in the process of star formation, has not been well studied observationally due to the short lifetime of the FHSC stage, difficulties in establishing the relative evolutionary stage of objects, the small spatial scales involved, and the dearth of strong molecular tracers of cores at high densities. Even with ALMA, dense condensations within starless cores are very rarely detected in large surveys \citep{dunham_2016,kirk_2017}, suggesting most cores are not dynamically evolved. If the dense core lifetime at different densities is similar to the free-fall time, for example, only a few cores should be at this evolutionary stage in a star-forming region like Ophiuchus \citep{kirk_2017}.  

\begin{figure*}[t]
\begin{center}
\includegraphics[width=\textwidth]{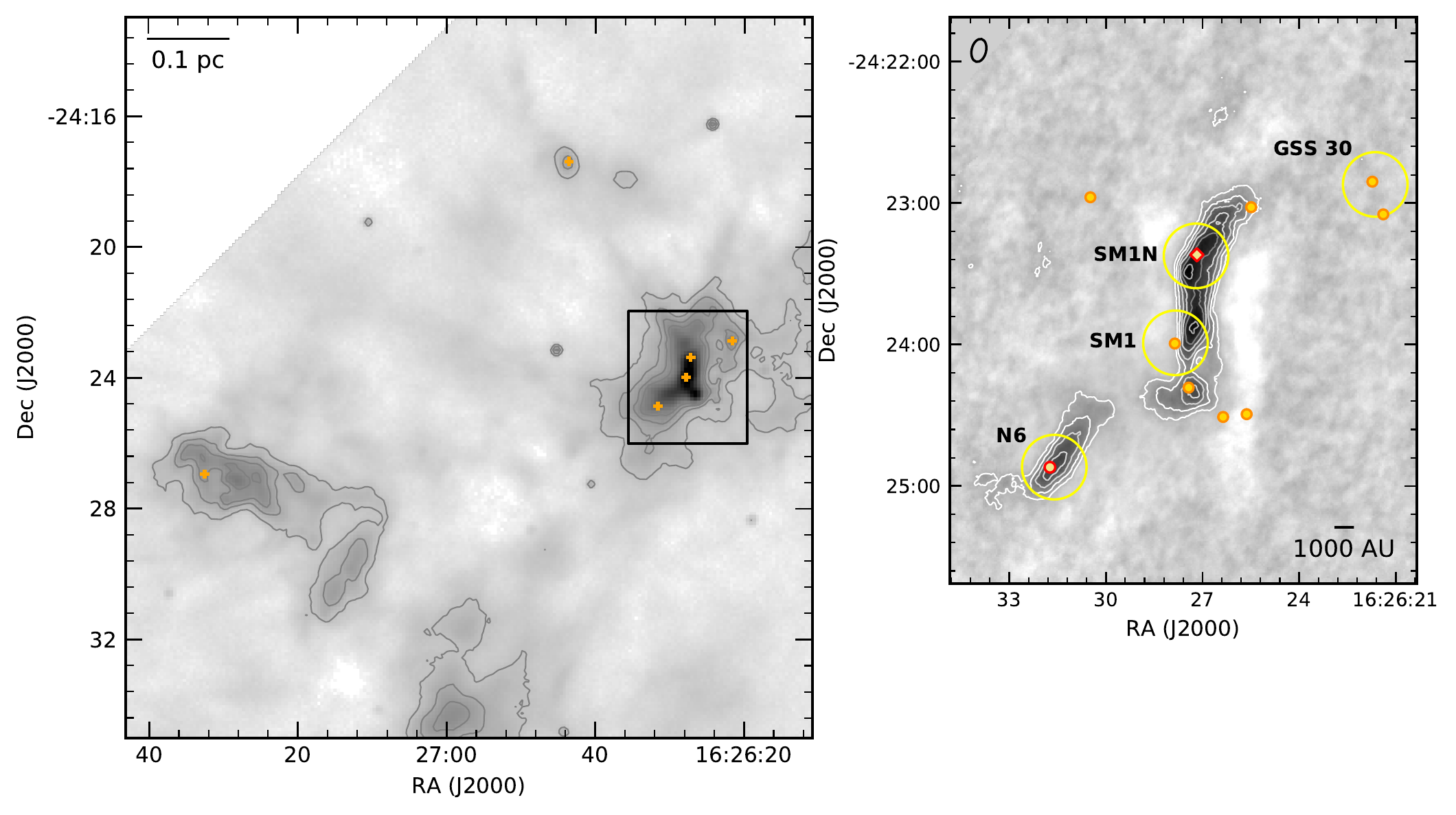}
\caption{Left: 850 \micron\ continuum emission toward part of L1688 in the Ophiuchus molecular cloud \citep[15\arcsec\ FWHM;][]{pattle_2015}. Gray contours are in intervals of 0.025 Jy\ pixel$^{-1}$. Orange crosses show the center pointing locations for the 6 ALMA targets. The black box highlights the Oph A region shown at right in more detail. Right: \dia\ 1-0 integrated intensity toward the Ophiuchus A ridge \citep[9\farcs9 $\times$ 6\farcs1 FWHM, shown at upper left;][]{difrancesco_2004}. White and gray contours highlight the \dia\ 1-0 integrated intensity. Yellow circles show the ALMA 12\ m array primary beam at 221\ GHz centred on four of the six sources presented here. Small orange circles show the locations of compact (sub)millimeter sources identified at 3\ mm with ALMA \citep{kirk_2017}, which include the well-known VLA 1623 and VLA 1623 W protostars (lower right), as well as sources like SM1 that were first detected by ALMA \citep{friesen14}, and have no known far-infrared counterparts. The red circle shows the location of the compact source toward N6, first detected unambiguously here, while the red diamond identifies the (sub)millimeter emission peak toward SM1N. \label{fig:opha_bima}}
\end{center}
\end{figure*}

\begin{figure*}[t]
\begin{center}
\includegraphics[width=0.98\textwidth]{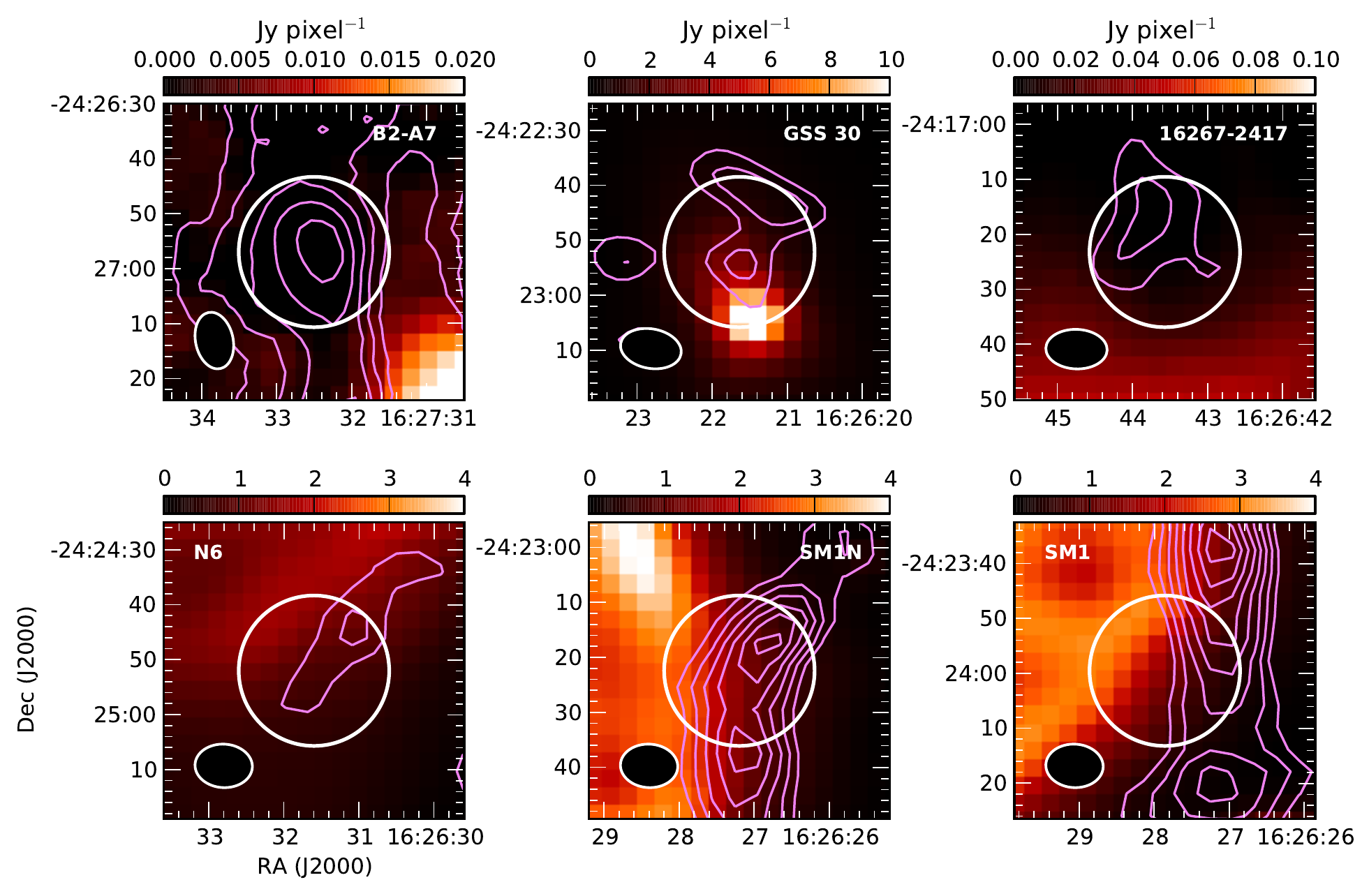}
\caption{70\ \micron\ continuum emission detected by \textit{Herschel} (colour scale; 5\farcs2 FWHM) toward the six targets in this study, with the images centered on the ALMA pointing coordinates. The white circle shows the ALMA primary beam at 1.1\ mm. Purple contours show \amm\ (1,1) emission at \revise{$\sim 10$\arcsec~$\times$~8\arcsec\ resolution (synthesized beam FWHM; shown at lower left in each panel)}, with levels beginning at $\sim 3\times\sigma$ and increasing by $2\times\sigma$.  \label{fig:herschel}
}
\end{center}
\end{figure*}

The Ophiuchus molecular cloud is our nearest example of active, \revise{low-mass, star formation} \citep[$d = 137.3 \pm 1.2$~pc \revise{via radio VLBA observations of young stars;}][\revise{see also \citealt{gagne_2018} for Gaia results giving $d=131\pm3$~pc}]{ortizleon_2017}. Within the most active star-forming region, L1688, the $\sim 30$ M$_\odot$ \revise{(gas and dust mass)} Oph A ridge contains a string of dense molecular cores that straddle the transition from prestellar to protostellar cores \citep[shown in Figure \ref{fig:opha_bima};][]{pon_2009,bourke12,murillo13a,friesen14}. 

In ALMA's Cycle 0, we observed two cores that we argued bracket the instant of star formation, SM1 and SM1N in Ophiuchus \citep[][hereafter \papi]{friesen14}. At high resolution, SM1 contains a resolved, likely protostellar source, where compact submillimeter continuum dust emission subtends only $\sim 37$\ au in size (Gaussian width). Although the collapse timescale for SM1 is $\lesssim 1000$~yr, the compact source is too large by a factor $\sim 2$ \revise{than the largest predictions from current} FHSC models \revise{(which require rapid rotation of the core)}.
SM1N contains a similar total mass over a larger, elongated structure. The mass and size of SM1N are consistent with either a highly concentrated starless core, or the first detection of a pseudodisk. Toward SM1N, emission from ortho-\hd\, was detected, the first such ALMA detection. The line is asymmetric and shows a velocity gradient of $\sim 0.1$~\kms\, over $\sim 400$\ au. The \hd\, abundance drops by a factor of a few toward the continuum peak, which could be due to the depletion of the parent molecule HD from the gas phase at a timescale $\gtrsim 2 \times 10^5$~yr \citep{sipila13}, or through heating by an extremely young protostar \citep{furuya12}. 

Here, we present ALMA observations of 1.1\ mm continuum and CO 2-1 emission, as well as \amm\ (1,1) observations with the Australia Telescope Compact Array (ATCA) toward SM1 and SM1N, along with several other highly-evolved starless and star-forming cores in the Ophiuchus molecular cloud. We find several candidate cores that are on the cusp of collapse and shed light on the formation of a FHSC or protostar. We describe the source selection and observations in Section \ref{sec:obs}. We present maps of the continuum emission and CO integrated intensity in Section \ref{sec:results}, and examine the source masses and dust opacities. We summarize the results and discuss the likely evolutionary status of the targets in Section \ref{sec:disc}. 

\section{Observations}
\label{sec:obs}

\subsection{Targets}
\label{sec:sources}

\floattable
\begin{deluxetable}{lccccccc}
\tablecolumns{8}
\tablewidth{0pt}
\tablecaption{Starless or very low luminosity protostellar core targets in Ophiuchus \label{tab:targets}}
\tablehead{
\colhead{Source} & \colhead{R. A.} & \colhead{decl.} & \colhead{\revise{$F_\mathrm{850}$}\tablenotemark{a}} & \colhead{$M_\mathrm{core}$
\tablenotemark{a}} & 
\colhead{Conc.\tablenotemark{b}} & \colhead{$T_K$ (GBT)\tablenotemark{c}} & \colhead{$T_K$ (ATCA)\tablenotemark{d}}\\
\colhead{} & \colhead{(J2000)} & \colhead{(J2000)} & \colhead{\revise{(Jy)}} & \colhead{(M$_\odot$)} & \colhead{} 
& \colhead{(K)} & \colhead{(K)} 
}
\startdata
GSS 30 	    & 16 26 21.64  & -24 22 52.1  & \revise{0.9} & \revise{0.1}\tablenotemark{e} & 0.55  & 23.9 (0.06) & \nodata \\
SM1N 		& 16 26 27.20  & -24 23 22.4  & \revise{6.8} & \revise{1.0} & 0.66 & 17.7 (0.003) & 15.5 (0.1)\\
SM1 		& 16 26 27.86  & -24 23 59.5  & \revise{5.2} & \revise{1.3} & 0.80 & 18.0 (0.003) & 24.2 (0.2)\\
N6 			& 16 26 31.60  & -24 24 52.0  & \revise{1.6} & \revise{0.3} & \nodata & 18.2 (0.01) & 16.5 (0.3) \\
16267-2417 	& 16 26 43.57  & -24 17 23.2  & \revise{1.2} & \revise{0.3} & 0.40 & \nodata & \nodata \\
B2-A7 		& 16 27 32.50  & -24 26 57.0  & \revise{0.9} & \revise{0.3} & \nodata & 13.8 (0.01) & 11.9 (0.1)\tablenotemark{f}\\
\enddata
\tablenotetext{a}{Based on continuum measurements at 14\arcsec\ resolution. \revise{Dust temperatures used for the mass calculations were derived from SED fits to the continuum emission toward each source \citep{pattle_2015}.}}
\tablenotetext{b}{Core concentration, determined from 850~\micron\ emission at 15\arcsec\ resolution where available \citep{johnstone00,jorgensen_2008}}
\tablenotetext{c}{Gas temperatures from \amm\ hyperfine line modeling at 32\arcsec\ FWHM. Uncertainties are the error in the weighted mean within a beam; typical uncertainties per pixel are a few $\times 0.1$\ K \citep{GAS_2017}.}
\tablenotetext{d}{Gas temperatures from \amm\ hyperfine line modeling at $\sim 9$\arcsec\ FWHM. Uncertainties are the error in the weighted mean within a beam; typical uncertainties per pixel are $\sim 0.5 - 1$\ K. }
\tablenotemark{e}{\revise{Masses based on single-dish continuum measurements for GSS 30 range from 0.1~\msun\ to 3.0~\msun\ \citep{johnstone00,jorgensen09,pattle_2015}.}}
\tablenotetext{f}{Gas temperature from \amm\ hyperfine line modeling at 15\arcsec\ FWHM \citep{friesen_2009}.}
\end{deluxetable}
\twocolumngrid

For this study, we identified cores in the Ophiuchus molecular cloud that are likely either on the cusp of star formation (identified by high core concentration, or the presence of compact structures observed in \amm\ emission with the ATCA, described below), or contain low luminosity, young protostellar objects. Table \ref{tab:targets} lists the position, submillimeter continuum flux, core mass and concentration \citep{jorgensen_2008,johnstone00}. Four sources lie within the Ophiuchus A ridge and B2 clump. The remaining two targets are more isolated from these larger, dense structures, but are nevertheless located within the central star-forming region, L1688, in Ophiuchus. The locations of all observed sources are shown in Figure \ref{fig:opha_bima}. 

The three targets within Oph A have known compact structure based on previous ALMA (\papi, SM1 and SM1N) and SMA \citep[][N6]{bourke12} (sub)millimeter continuum observations. SM1 and SM1N are described in more detail above, and in \papi.

Oph A N6 (hereafter N6) was identified as one of eight maxima in \dia~1-0 emission along the Oph A core by \citet{difrancesco_2004}. At the high densities traced by \dia\ ($\gtrsim 2 \times 10^5$\ \cc), N6 is an elongated structure with very narrow line widths ($\Delta v \sim 0.19$\ \kms) and mass $M \sim 0.29$\ M$_\odot$ \citep{bourke12}. It is abundant in deuterated molecules, typical of evolved starless cores, and optically thick spectral lines reveal evidence for infall or collapse \citep{pon_2009,bourke12}. At 4\farcs6 $\times$ 3\farcs5 resolution, \citeauthor{bourke12} detect a compact millimeter continuum source with size $\sim 1000$~au and mass $M \sim 0.005 - 0.01$~M$_\odot$. No mid-infrared point source, down to 0.001~L$_\odot$, was detected toward N6 by the Spitzer Cores to Disks (c2d) survey \citep{evans_2003, padgett_2008,dunham_2015}.

Oph B2-A7 was identified in \amm\ (1,1) emission (15\arcsec\ resolution) as a compact structure embedded within the Ophiuchus B2 core with mass $M \sim 0.2$\ M$_\odot$, that is furthermore cold with subsonic velocity dispersion \citep[$\sigma_v \sim 0.2$~\kms;][]{friesen_2009}. B2-A7 has a high mean density \citep[$n \gtrsim 10^5 - 10^8$~\cc;][]{friesen_2009,motte98}, and despite its low mass it is more massive than a critical Bonnor-Ebert sphere at the observed temperature \citep{motte98}. 

One of the two more isolated targets, GSS 30, is a \revise{low-mass} core that is associated with three infrared sources \citep[GSS 30 IRS1, GSS 30 IRS2, GSS 30 IRS3;][]{grasdalen_1973,elias_1978,weintraub_1993}. GSS 30 IRS2 is a weak-lined T Tauri star \citep{DoAr_1959} not embedded within the dense core. Offset by $\sim 15$\arcsec\ from the 850~\micron\ continuum emission peak, GSS 30 IRS1 is a Class I protostar with $L_{\mathrm{bol}} \sim 11-18$\ L$_\odot$ and $T_\mathrm{bol} \sim 133-172$~K \citep{enoch_2009,green_2013}. GSS 30 IRS3 is also known as the 6\ cm radio continuum source LFAM1 \citep{leous91}, and was classified as a Class I protostar with an estimated $L_\mathrm{bol} \sim 0.13$~L$_\odot$ based on a faint 6.7~\micron\ detection \citep[][]{bontemps_2001}, but it is not identified as a protostar in the c2d survey \citep{dunham_2015}. \citet{jorgensen09} note that IRS1 and IRS3 are not separated in the Spitzer catalog, with IRS1 dominating the infrared emission, possibly confusing the $L_{\mathrm{bol}}$ and $T_\mathrm{bol}$ measurements. Assuming this luminosity, however, GSS 30 IRS3 is a borderline VeLLO. With the Owens Valley Radio Observatory Millimeter Array, \citet[][8\farcs9 $\times$ 4\farcs6 angular resolution]{zhang_1997} detected an unresolved millimeter source coincident with GSS 30 IRS3. Observations of CO failed to detect any outflow associated with the source. With the SMA, \citet{jorgensen09} detect compact line emission but no 1.3~mm continuum emission associated with IRS1, and continuum emission but no compact line emission toward IRS3. 

In contrast to GSS 30, 16267-2417 is a dense core that is moderately-to-significantly peaked in 850~\micron\ continuum emission at 15\arcsec\ resolution \citep{jorgensen_2008,johnstone01}, suggesting its density profile is steep. It has no known infrared or radio counterpart.

 We show in Figure \ref{fig:herschel} the 70~\micron\ continuum emission \citep[5\farcs2 FWHM, \textit{Herschel};][]{alves_do_2013} toward the six targets, which highlights warm dust and embedded protostars. None of the targeted cores contain 70~\micron\ point source emission within 15\arcsec\ of the core centre. Toward GSS 30, a 70~\micron\ point source corresponding to GSS IRS1 is seen offset just over 15\arcsec\ to the south from the core centre. Furthermore, the proximity of most of the targets to bright, extended 70~\micron\ emission (particularly toward the Oph A ridge) precludes determining robust upper limits on the internal luminosity of an embedded protostar within the cores \citep[e.g., using the tight correlation between \lint\ and the 70\ \micron\ luminosity of a protostar; see][]{dunham08}. \revise{Similarly, no clear detections are seen toward any targets in 100~\micron\ Herschel data.}

In \papi, we assumed a distance of $120\pm 4$~pc toward Ophiuchus as derived through parallax measurements of two young stars by \citet{loinard08}. New parallax measurements of 26 stellar systems put the central \revise{region of highest star formation activity}, L1688, at a slightly further distance of $137 \pm 1.2$~pc \citep{ortizleon_2017}. In this paper we use this updated distance. 

\subsection{Australia Telescope Compact Array}
\label{sec:atca}

The Ophiuchus A region, containing the SM1, SM1N, and N6 sources, was observed in emission from the \amm\ (1,1) and (2,2) inversion transitions with the ATCA between July 2010 and August 2010 (project ID C2322; PI R. Friesen). A mosaic of five pointings in a hexagonal pattern was used to cover the main ridge and narrow extension to the south-east as seen in interferometer observations of \dia\ \citep{difrancesco_2004}. Flux and bandpass calibrations were performed each observing shift on the standard calibrators 1934-638 and 1253-055, while phase calibration was performed every 20 minutes on a nearby quasar (1622-297; 5.3\arcdeg\ from the mosaic centre). Observations cycled through each mosaic pointing four times between phase calibrator observations to ensure broad coverage of the uv-plane. The total on-source time per mosaic pointing was $\sim 180$ minutes; however due to poor weather and noise issues, substantial data had to be flagged, with the resulting rms noise in the final cubes much greater than expected for the given time on-source. 

Calibration was performed using the miriad software package \citep{sault95}. The synthesized beam of the mosaic is 10\farcs6 $\times$ 7\farcs0, with a position angle of 87\arcdeg. 

Single pointing observations of \amm\ (1,1) and (2,2) emission from the isolated cores GSS 30 and 16267-2417 in the Ophiuchus cloud were performed with the ATCA in August 2006 (project ID C1566; PI R. Friesen). The same calibrator targets were used for flux, bandpass and phase calibration as in the 2010 observations, with the same 20 minute cadence for phase calibration observations. Each target was observed for two 20 minute intervals. The lack of cycling between targets resulted in less coverage of the uv-plane for individual pointings than in the 2010 observations. Consequently, the sidelobe amplitudes in the synthesized beam are substantial, and the sidelobes were not well removed in the deconvolution process. Here, we primarily use the integrated intensity maps of the \amm\ (1,1) data to identify where compact, dense gas is present within the core at $\sim 9$\arcsec. 

\begin{figure}
\begin{center}
\includegraphics[width=0.49\textwidth]{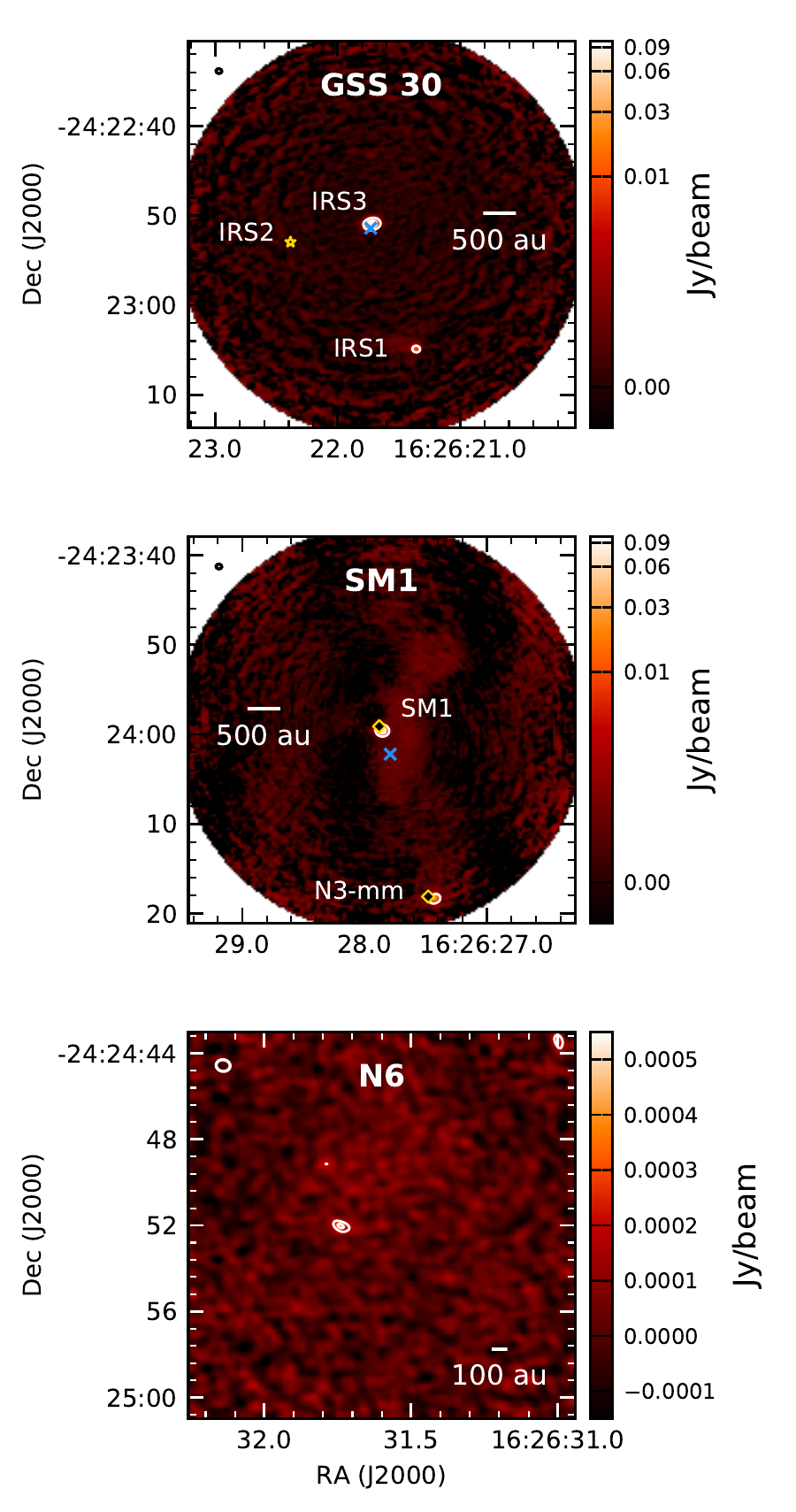}
\caption{1.1\ mm continuum ALMA images of the three targets with detected compact emission in the untapered ALMA data. Contours for GSS 30 and SM1 are [50, 500]$\times \sigma$, where $\sigma = 0.08$~mJy~beam$^{-1}$ and $\sigma = 0.11$~mJy~beam$^{-1}$, respectively. Compact radio (blue `x') and x-ray (yellow diamond) detections are identified \citep{leous91,gagne04}. \revise{Toward GSS 30, we show the location of IRS2 (yellow star), which is undetected in the continuum.} Contours for N6 are [4, 6]$\times \sigma$, where $\sigma = 0.04$~mJy~beam$^{-1}$. A single compact source with S/N $\sim 7$ is detected. Note that the imaged area for N6 is smaller than that of SM1 and GSS 30. For all sources, the 0\farcs66 $\times$ 0\farcs55 synthesized beam is shown at upper left. Scale bars are also shown. \label{fig:cont_notaper}}
\end{center}
\end{figure}

Lastly, the starless core B2-A7 is within the Ophiuchus B region. Oph B was observed in \amm\ (1,1) and (2,2) emission with the ATCA, the Very Large Array (VLA) and the Green Bank Telescope (GBT), with the data tapered to produce a synthesized beam of 15\arcsec\ for the combined data to improve the S/N \citep{friesen_2009}. Here, we use the combined \amm\ data at the untapered resolution, with a synthesized beam of 10\farcs6 $\times$ 8\farcs4 at a position angle of 12\arcdeg. 

In Figure \ref{fig:herschel}, purple contours show the integrated intensity of \amm\ (1,1) emission toward the six footprints. 

\subsection{Atacama Large Millimeter/submillimeter Array}
\label{sec:alma}

The ALMA data presented here were obtained as part of the Cycle 2 project 2013.1.00937.S (PI: R. Friesen). Each of the six targets was observed in 1.1\ mm continuum emission and emission from the CO $J=2-1$ rotational transition (230.5 GHz) in single pointings of the ALMA 12m Array on 24-May-2015 and 25-May-2015. The total observation time per target was approximately 15\ minutes. Thirty-seven antennas were available for these observations, giving a total of 666 baselines that ranged in length from 21~m to 555~m. 

Three continuum spectral windows were used, each with 2~GHz bandwidth, at 214.5~GHz, 217.2~GHz, and 227.8~GHz, giving a total combined bandwidth of 6~GHz. The CO spectral window, centered at 230.3~GHz, had a bandwidth of 468.8~MHz and a spectral resolution of 122~kHz (0.125~\kms). At 1.1~mm, the primary beam of the 12m Array is 27\farcs4. The largest angular scale recoverable given the observing frequency and the minimum baseline length is $\sim 6$\farcs4. 

The calibration and imaging were done using the Common Astronomy Software Applications (CASA) package \citep{CASA_2007}. Nearby quasars were used as phase calibrators, and flux calibration was performed using Titan assuming the Butler-JPL-Horizons 2012 flux model, which is
expected to be accurate to within 15\%\footnote{\url{https://library.nrao.edu/public/memos/alma/memo594.pdf}}. 

Continuum images for each target were made with the \textit{clean} task in CASA, using multifrequency synthesis to combine data from the three continuum spectral windows. Using a robust weighting of 0.5, the resulting synthesized beam is roughly 0.6\arcsec ($\sim 80$~au). For several sources, the continuum data were tapered to enhance the SNR, with a resulting synthesized beam of $\sim 1$\farcs5 (205~au). Self-calibration was performed on SM1 and GSS 30, using the detected bright compact sources as the initial model, to reduce the rms noise and improve the dynamic range. In all cases, the data were deconvolved using multiscale clean with scales of 0, 5, and 10 times the synthesized beam and corrected for the primary beam. The resulting rms noise in the continuum images ranges from 40~\muJy~beam$^{-1}$ to 100~\muJy~beam$^{-1}$. In the tapered images, the 1$\times\sigma$ rms noise is $\sim 70$~\muJy~beam$^{-1}$. 

We next imaged the CO emission. For SM1 and GSS 30, the self-calibration solutions derived from the continuum were applied to the line spectral window. For all sources, the continuum emission (measured in the line-free spectral channels) was subtracted from the CO spectral window in the uv-plane. The continuum-subtracted visibilities were then imaged with the same robust and multiscale parameters used for the continuum images,  a velocity resolution of 0.25~\kms, and were corrected for the primary beam. For all sources, the CO emission is heavily self-absorbed, with little to no flux at the typical local standard of rest velocity for L1688 in Ophiuchus, \vlsr\ $\sim 2.5$~\kms. We thus focus mainly on the line wings of the CO emission in the following discussion, but also discuss the strong, compact CO self-absorption seen toward the SM1 and GSS 30 continuum sources.  

\section{Results and Analysis}
\label{sec:results}

\subsection{Millimeter continuum detections} 

We detect 1.1~mm continuum emission with SNR $\gtrsim 5$ toward five of the six targets, either at the original resolution of the data, or in the lower-resolution, tapered images as described above. The 16267-2417 core is the exception, with no emission detected to the sensitivity of the ALMA observations.  

\begin{figure}
\begin{center}
\includegraphics[width=0.95\columnwidth,trim=5 0 0 0, clip]{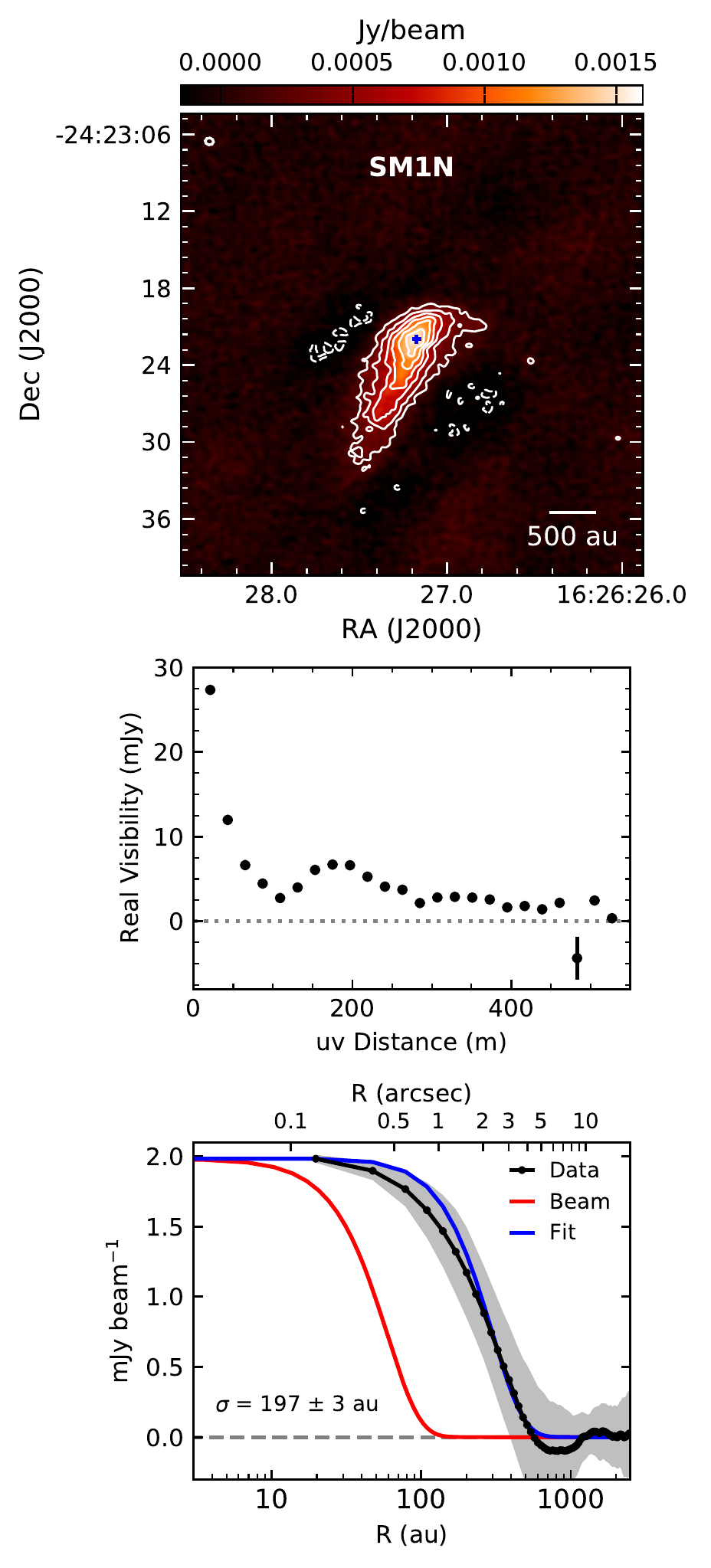}
\caption{Top: 1.1~mm continuum ALMA image of SM1N, uncorrected for the primary beam. Contours are [5, 10, 15, ...]$\times\sigma$, where $\sigma = 0.05$~mJy~beam$^{-1}$. The 0\farcs66 $\times$ 0\farcs55 synthesized beam is shown at upper left. \revise{The emission peak is identified by the blue cross.}
\revise{
Middle: Binned real visibilities toward SM1N. Error bars (where visible) show the standard deviation in each bin. The phase center was updated to match the emission peak, and the uv-data were corrected using the \texttt{fixvis} task in CASA. 
Bottom: Radial profile of SM1N derived from the primary beam-corrected continuum image, centered on the emission peak. The synthesized beam profile is shown in red, while the best-fit Gaussian profile convolved with the beam is shown in blue. Grayscale shows one standard deviation in each radial bin. }
\label{fig:sm1n_profile}}
\end{center}
\end{figure}

\subsubsection{Compact sources}

Three of the sources (GSS 30, SM1, and N6) each show either one or two compact, sub-arcsecond continuum structures. Figure \ref{fig:cont_notaper} shows the continuum emission toward GSS 30 (top), SM1 (middle), and N6 (bottom). Table \ref{tab:gaussfit} lists the source names and peak locations of all detected compact sources. 

Toward GSS 30, we detect GSS 30 IRS3 at a SNR $\gtrsim 500$ in the centre of the primary beam, as well as GSS 30 IRS1 at a SNR $\sim 50$. The peak locations of the compact continuum emission agree well with previous identifications of these sources. Both sources were also detected in 3~mm ALMA observations at lower angular resolution \citep[3\farcs5 $\times$ 1\farcs8 FWHM;][]{kirk_2017}. No extended emission is visible either in our image or uv-data. The emission from the extended dense core appears entirely resolved out. 

Toward SM1, we detect two bright, compact continuum sources, with similar flux densities and SNR as toward GSS 30. The central, compact source in SM1 was previously identified in \papi, while the second component has previously only been detected at millimeter/submillimeter wavelengths with ALMA at 3~mm \citep[][respectively source 10 and Source X]{kirk_2017,kawabe_2018}. The secondary source location agrees within 3\farcs5 the position of a \dia~1-0 peak identified as N3 \citep{difrancesco_2004}. Going forward, we thus label this peak N3-mm. We also detect faint extended continuum emission up to the maximum angular scale measured by the ALMA observations in the pointing that is aligned with the larger-scale Ophiuchus A ridge again traced by \dia\ emission. 

Point-source 6~cm radio emission has been detected toward GSS 30 IRS3 and SM1 \citep[][identified with a blue 'x' in Figure \ref{fig:cont_notaper}]{leous91,gagne04}. GSS 30 IRS1 and N3-mm do not have recorded radio detections. Both SM1 and N3-mm are additionally coincident with variable hard x-ray sources \citep[][green crosses in Figure \ref{fig:cont_notaper}]{gagne04}, indicative of young, actively accreting protostars.

Toward N6, a single compact source is detected with a SNR $\sim 7$. The compact source detected toward N6 is offset by only 1\farcs8 from the observed \dia~1-0 integrated intensity emission peak identified by \citet[][angular resolution of 9\farcs9 $\times$ 6\farcs1 FWHM]{difrancesco_2004}, and matches well the 1.1~mm continuum emission peak location observed with the SMA \citep[][angular resolution of 4\farcs6 $\times$ 3\farcs5 FWHM]{bourke12}. We thus label this source N6-mm. 

For the five detected compact sources, we fit the continuum image with one or two 2D Gaussians using the CASA task \texttt{imfit}. We list in Table \ref{tab:gaussfit} the fit results. The sources thus identified range in angular width along their major axis from being unresolved to 1\farcs3, or $\lesssim 40$~au to $\sim 180$~au at $d = 137.3$~pc. 
Three of the four resolved sources are elongated, with aspect ratios $\geq 2$ (GSS 30 IRS3, N3-mm, and N6). These values are at the high end of the $\sim 1-2$ aspect ratios typically found for star-forming cores. In contrast, SM1 has an aspect ratio of only 1.1. We note that for GSS 30 IRS3 and SM1, we also fit directly the observed visibilities with both a disk and Gaussian model using the CASA task \texttt{uvmodelfit}, and found that both geometries fit the uv-data equally well as measured by the reduced $\chi^2$ between the data and model. 

\begin{figure}[t]
\includegraphics[width=0.99\columnwidth,trim=5 0 5 5,clip]{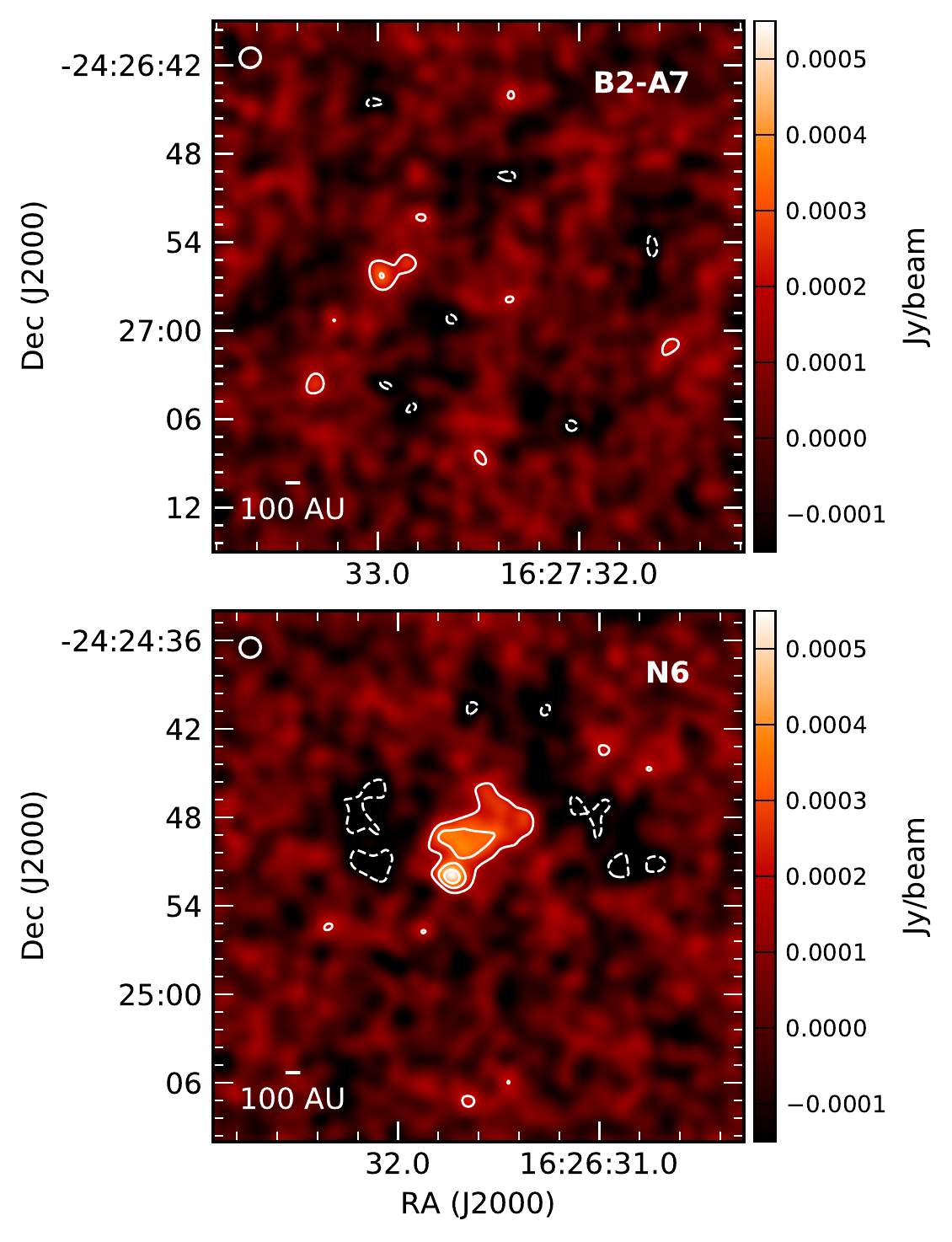}
\caption{1.1~mm continuum ALMA images of B2-A7 (top) and Oph A N6 (bottom), tapered to a synthesized beam of 1\farcs42 $\times$ 1\farcs36 (shown at upper left) with rms noise of 0.06~mJy~beam$^{-1}$. The images have not been corrected for the ALMA primary beam, and are approximately the width of the primary beam (27\farcs4 FWHM). In both panels, solid white contours highlight [3, 5, 7, 9]$\times \sigma$ in the continuum image, and dashed contours show negative bowls in the image at [-3,-5]$\times \sigma$. \label{fig:cont_taper}}
\end{figure}

\floattable 
\begin{deluxetable}{lccccccccccc} 
\tabletypesize{\footnotesize} 
\tablecolumns{12} 
\tablewidth{0pt} 
\tablecaption{Gaussian fit results for compact sources \label{tab:gaussfit}} 
\tablehead{ 
\colhead{Source} & \colhead{R.A.} & \colhead{decl.} & \colhead{$S_\nu$} & \colhead{$\sigma_\mathrm{maj}$} & \colhead{$\sigma_\mathrm{min}$} & \colhead{P.A.} & \colhead{$M$\tablenotemark{a}} & \colhead{$n$\tablenotemark{a}} & \colhead{$\beta$} & \colhead{$M$\tablenotemark{b}} & \colhead{$n$\tablenotemark{b}} \\ 
\colhead{} & \colhead{J2000} & \colhead{J2000} & \colhead{Jy} & \colhead{\arcsec} & \colhead{\arcsec} & \colhead{\degr} & \colhead{M$_\odot$} & \colhead{cm$^{-3}$} & \colhead{} & \colhead{M$_\odot$} & \colhead{cm$^{-3}$} 
} 
\startdata 
GSS 30 IRS1 & 16:26:21.359 & -24:23:04.865 & 0.0136(8) &  $\leq 0.41$ &  $\leq 0.19$ &  \nodata &  0.0043 & $\geq 3.4$e+09 & 0.0(5) & 0.0005 & $\geq 4.0$e+08 \\ 
GSS 30 IRS3 & 16:26:21.720 & -24:22:50.939 & 0.1423(10) &  0.506(  6) &  0.169( 15) &  0.0(1.7) &  0.0446 & 4.3e+10 & 0.2(2) & 0.0046 & 4.5e+09 \\ 
N3-mm & 16:26:27.428 & -24:24:18.317 & 0.0416(9) &  0.56(2) &  0.275(32) &  125.9(4.3) &  0.0130 & 4.4e+09 & 0.5(2) & 0.0021 & 7.1e+08 \\ 
SM1 & 16:26:27.856 & -24:23:59.570 & 0.1289(8) &  0.321(6) &  0.297(9) &  179(18) &  0.0404 & 2.0e+10 & 0.2(2) & 0.0042 & 2.1e+09 \\ 
N6-mm & 16:26:31.732 & -24:24:52.044 & 0.0007(1) &  1.3(4) &  0.4(3) &  78.4(9.6) &  0.0002 & 1.5e+07 & \nodata & \nodata & \nodata \\ 
\enddata 
\tablecomments{All sizes and position angles have been deconvolved with the synthesized beam.}
\tablenotetext{a}{Masses determined with $\beta = 1.7$.} 
\tablenotetext{b}{Masses determined with $\beta$ derived in \ref{sec:masses}.} 
\end{deluxetable}

\subsubsection{Extended structure}

In Figure \ref{fig:sm1n_profile} (top) we show the continuum emission toward SM1N. The observed structure is extended along a northwest-to-southeast axis, and contains a compact central emission peak towards the northwest end. This extension is also seen at 3~mm, with hints of a second, fainter peak in the southeast \citep{kamazaki03,kirk_2017}. Our non-detection here of this peak may be due to the sensitivity drop near the edge of the primary beam, as well as the maximum recoverable scale of these observations. 

Strong evidence for compact emission is also seen in the uv-data, shown in Figure \ref{fig:sm1n_profile} (middle), where the real visibilities consistently remain above zero out to the maximum baselines of the ALMA data. The structure in the uv-data suggests the superposition of two extended sources, or a resolved disk. Consequently, the emission toward SM1N is poorly fit by a single Gaussian in either the image or uv-plane. In addition, we are unable to find an unique solution for a combination of compact and extended emission. Instead, we show in Figure \ref{fig:sm1n_profile} (bottom) the radially-averaged continuum emission (black curve and points) centered at the position of brightest flux density, identified with the blue cross. The red curve represents a Gaussian profile with the effective width of the synthesized beam. We fit the observed intensity profile with a single Gaussian, convolved with the synthesized beam, shown as the blue curve in the Figure. The best fit Gaussian has a width \revise{$\sigma = 1.44 \pm 0.02$\arcsec, or $197 \pm 3$~au} (blue curve). A comparison of the observed radial profile and model fit, however, shows that at radii \revise{$< 1.5$\arcsec\ ($< 200$~au), the radial profile} is more peaked than predicted by the best-fit Gaussian profile.

In Figure \ref{fig:cont_taper} we show the tapered continuum images for N6 and B2-A7. Toward N6, the compact source seen in Figure \ref{fig:cont_notaper} remains with a SNR $\sim 10$, and extended emission is also visible with SNR $\sim 3-5$. The extended emission observed with ALMA agrees well with previous 1.3~mm continuum observations of N6 with the SMA \citep[$\sim 4$\arcsec\ resolution;][]{bourke12}. Toward B2-A7, we find some diffuse emission in the tapered image with a peak SNR $\sim 5$. 

The extended features seen toward SM1N and N6 have extents similar to the largest recoverable angular scale of these observations ($\sim 6$\arcsec). These larger features may represent the observable peak of an underlying larger distribution of dense gas. 

\subsection{Masses of continuum condensations}
\label{sec:masses}

Given the total flux of each component, $S_\nu$, we calculate the mass $M$ following 

\begin{equation}
M = \frac{d^2 \ S_\nu}{\kappa_\nu B_\nu (T_d)}, 
\label{eqn:mass}
\end{equation}

\noindent where $d = 137.3$~pc is the source distance, $\kappa_\nu$ is the dust opacity at the frequency $\nu$ of the observations, and $B_\nu (T_d)$ is the Planck function at frequency $\nu$ and dust temperature $T_d$. The dust opacity $\kappa_\nu = \kappa_0 (\nu/\nu_0)^\beta$, where both the normalization $\kappa_0$ at frequency $\nu_0$ and the spectral index, $\beta$, are dependent on the assumed dust grain properties. Equation \ref{eqn:mass} assumes that the continuum emission is optically thin, an assumption we will review below. 

While we are unable to ascertain $T_d$ on 0\farcs5 - 1\arcsec\ scales here, we have precise measurements of the gas temperatures, \tkin, at 32\arcsec\ resolution \citep[from the Green Bank Ammonia Survey, or GAS;][]{GAS_2017} and $\sim 8-15$\arcsec\ from detailed \amm\ model fitting \citep[][and this paper]{friesen_2009}. At the densities expected for the ALMA-detected sources, the gas and dust are well-coupled \citep[$n \gtrsim 10^{4.5}$~\cc;][]{goldsmith_2001}. All sources but 16267-2417 have precise \tkin\ measurements from GAS. GSS 30 and 16267-2417 do not have high resolution \tkin\ measurements. We list in Table \ref{tab:targets} the uncertainty-weighted average \tkin\ values measured at low and high resolution toward each ALMA target, where the average is made over an area with effective radius $r_\mathrm{eff} = \sqrt{b_\mathrm{maj} b_\mathrm{min}}$ ($b_\mathrm{maj}$ and $b_\mathrm{min}$ are the major and minor beam widths). 

On larger scales, the dense gas in Ophiuchus is slightly warmer (\tkin\ $\sim 15-20$\ K) than generally measured toward less actively star-forming regions like Taurus \citep[$\sim 12$~K; ][]{ho_1977}. Relative to the lower resolution measurements, the sources without strong, compact continuum emission in the ALMA observations (N6, B2-A7, and SM1N) are cooler at higher resolution. In contrast, SM1 appears warmer at higher resolution. 

Accordingly, we use $T_d = 15$~K when calculating the mass of the extended emission seen by ALMA, and set $T_d = 30$~K for the compact sources, to better compare our results with previous studies of protostellar disk emission around Class 0/I protostars \citep{jorgensen07,jorgensen09}, including VLA 1623A in Oph A \citep{murillo13a}. We note, however, that this assumption likely underestimates the total mass of the disk, by up to factors of $\sim 2$ in Class I objects, as bulk temperatures across the disk are likely lower \citep{jorgensen09,dunham_2014}. 

For the dust opacity, we set $\kappa_0 = 0.1$~cm$^{2}$~g$^{-1}$ at a frequency $\nu_0 = 1000$~GHz. We initially assume $\beta=1.7$, a typical value used for dust in dense cores that is consistent with models of ice-coated dust grains \citep[the OH5 model;][]{ossen94}. The mass sensitivity in the ALMA beam is $\sim 2 \times 10^{-4}$~M$_\odot$ ($5\times\sigma$) given the 1$\times\sigma$ rms noise, the above dust parameters, and $T_d = 15$~K. 

\begin{figure}
\includegraphics[width=\columnwidth,trim=5 0 0 0,clip]{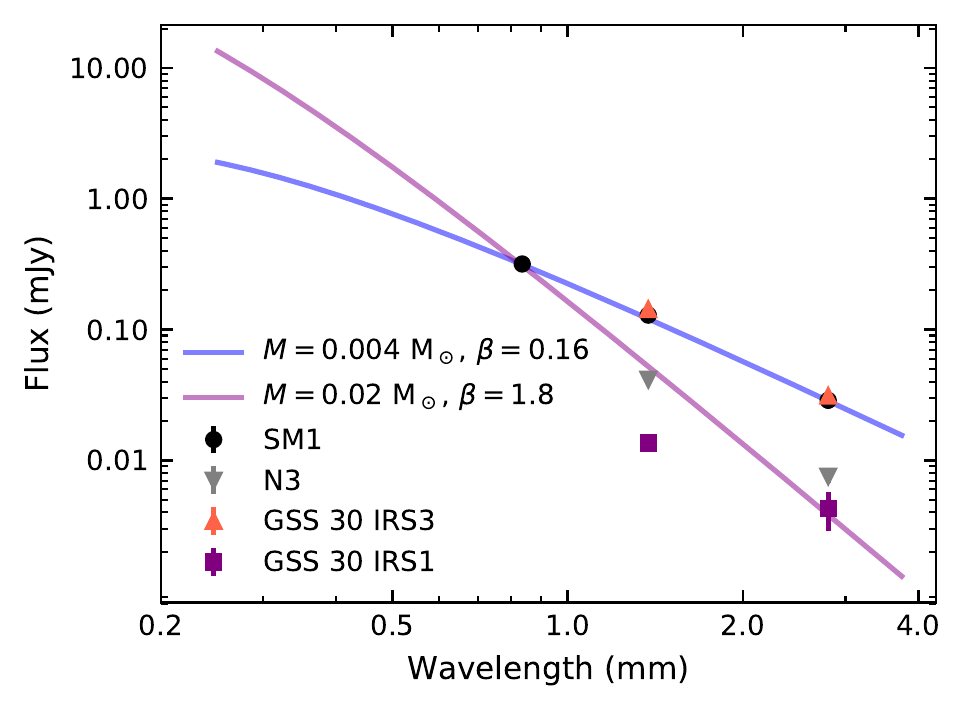}
\caption{Fluxes for compact sources detected by ALMA in this study, at 850~\micron\ \citep[SM1;][]{friesen14} and at 3~mm \citep{kirk_2017}, after convolving to match the 3~mm beam. For comparison, the expected SED corresponding to two different values of the dust spectral index, $\beta$, are shown, with masses chosen to match the SM1 850~\micron\ datapoint. None of the compact sources is consistent with $\beta = 1.8$, typically assumed for optically thin dense cores. \label{fig:beta}}
\end{figure}

\begin{figure}
\includegraphics[width=\columnwidth]{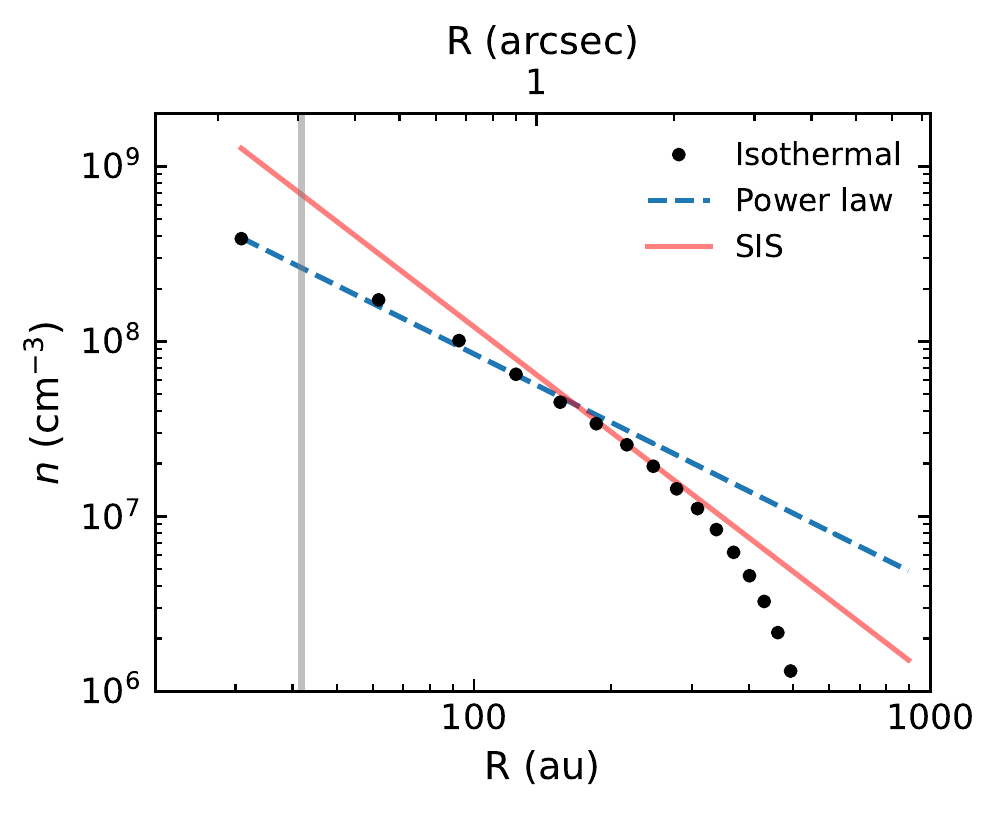}
\caption{Number density as a function of radius for SM1N in circular annuli centered at the emission peak identified in Figure \ref{fig:sm1n_profile}.  \revise{The vertical gray line shows the effective radius of the ALMA beam.} The blue line shows the best-fit power law \revise{to the data interior to $\sim 200$~au (1.5\arcsec)} with exponent -1.3. The red line shows the expected density profile for a singular isothermal sphere at a temperature of 15~K. \label{fig:SM1N_density}}
\end{figure}

In \papi, we calculated a lower value of $\beta = 0.4$ for SM1 via a Rayleigh-Jeans analysis of 227~GHz and 359~GHz continuum observations. Lower values of $\beta$ are suggestive of grain growth in disks, particularly at later stages \citep[e.g.,][]{beckwith_1991}, however significant grain growth has been argued for dust associated with some Class 0 protostars \citep{tobin_2013}. 

Without assuming that the emission at both 3~mm and 1~mm is on the Rayleigh-Jeans tail of the dust emission spectral energy distribution, we follow \citet{schnee_2014} to determine $\beta$ from the total flux of all four detected compact sources at these two wavelengths, assuming $T_d = 30$~K. We list the results in Table \ref{tab:gaussfit}. For three of the four compact sources we find $\beta \sim 0.2 - 0.5$. For GSS 30 IRS1, we find $\beta$ consistent with zero, but with larger uncertainty. Here, we assume a 10\% calibration error at both frequencies, and that the relative calibration is the largest source of error in the calculation given the small uncertainties in the total flux measurements. 

While these dust opacity values are consistent with grain growth in disks, the extremely low values determined here may also indicate that the optical depth of the continuum emission at wavelengths $\lesssim 1$\ mm is very high over part or all of the disks (in this analysis, the emission from a pure blackbody would be described by $\beta = 0$). Detailed analysis of the continuum emission of the more evolved HL Tau protoplanetary disk shows that the 1~mm optical depth $\tau \gtrsim 4.6$ in the detected dust rings out to $\gtrsim 100$\ au, while the rings within $\lesssim 50$~au are also optically thick at 3~mm \citep{pinte_2016}. Accurate, high resolution temperature measurements across the disks are needed to state definitively whether grain growth, optical depth, or some combination of the two is contributing to the lower $\beta$ values found here. In the optically thick regime, the derived masses are a lower limit \citep{evansm_2017}, although \citet{dunham_2014} argue that Class 0 and Class I disk mass underestimates due to optically thick emission are generally less than a factor of $\sim 2$ if the disk is not viewed edge-on.

We further list in Table \ref{tab:gaussfit} the average number density, $n = M/(\mu m_\mathrm{H} V)$, for each component, using both $\beta = 1.7$ and $\beta$ as derived above. We assume the sources are prolate spheroids such that the volume $V = 4/3 \ \pi \sigma_\mathrm{min}^2\sigma_\mathrm{maj}$. For GSS 30 IRS1, which is unresolved, we calculate the density upper limits. 

For SM1N, we find a total flux of $\sim 40$~mJy out to a radius of 3\arcsec. The total mass traced by the continuum emission in SM1N is therefore $\sim 0.03$~\msun, with a mean density $\sim 10^7$~\cc, assuming $T_d = 15$~K and $\beta = 1.7$. These values are consistent within uncertainties with those found in \papi\ at 359~GHz assuming similar dust parameters ($M \sim 0.02$~\msun). In contrast to SM1, this consistency suggests that the typical dense core dust properties describe well the observed emission from SM1N, and/or that the optical depth is not high toward SM1N. 

Figure \ref{fig:SM1N_density} shows the radial density profile of SM1N, calculated assuming spherical symmetry. Here, we have determined the total mass interior to each radial bin using Equation \ref{eqn:mass} and the dust properties above, then calculated the resulting density in concentric shells. We find a central density $n_c \sim 10^9$~\cc\ for SM1N, two orders of magnitude greater than the mean density $n \sim 10^7$~\cc. The Figure also shows the density profile is well fit by a single power law with exponent -1.3 at radii $\lesssim 1.5$\arcsec\ (\revise{$\lesssim 200$~au; fit }shown in blue), as suggested by the radial emission profile in Figure \ref{fig:sm1n_profile}. \revise{The density profile is similar to that found for compact, dense gas condensations in B5 \citep{pineda_2015}, and suggests that SM1N is out of hydrostatic equilibrium \citep{shu_1977}. We have overlaid the density profile of a singular isothermal sphere (SIS), where $\rho(r) = a^2 / (2 \pi\ G) r^{-2}$, and $a = (k_B T/m)$ is the thermal sound speed of the core gas \citep{shu_1977}. The SIS density profile} decreases more steeply as $1/r^2$, shown in red. SM1N is therefore not as centrally peaked as a SIS, but also does not show the flattened density profile at small radii expected for an isothermal Bonnor-Ebert sphere. We note, however, that if a compact, low-luminosity central source exists within SM1N such that the innermost few tens of au are heated to $\sim 30$~K, the true density profile would be flattened at \revise{small} radii. Conversely, if the core has a decreasing temperature profile with decreasing radius, the density profile would peak at higher values. Kinematics from dense gas tracers are needed to distinguish between stable and collapsing cores \citep{keto_2010,keto_2015}, and future ALMA observations will investigate further the stability of SM1N. 

Toward N6, we find a similar mass ($M \sim 0.025$\ M$_\odot$), but lower mean density ($n \sim 7 \times 10^5$\ \cc) for the extended emission in the tapered continuum image. 

\begin{figure}
\begin{center}
\includegraphics[width=0.99\columnwidth]{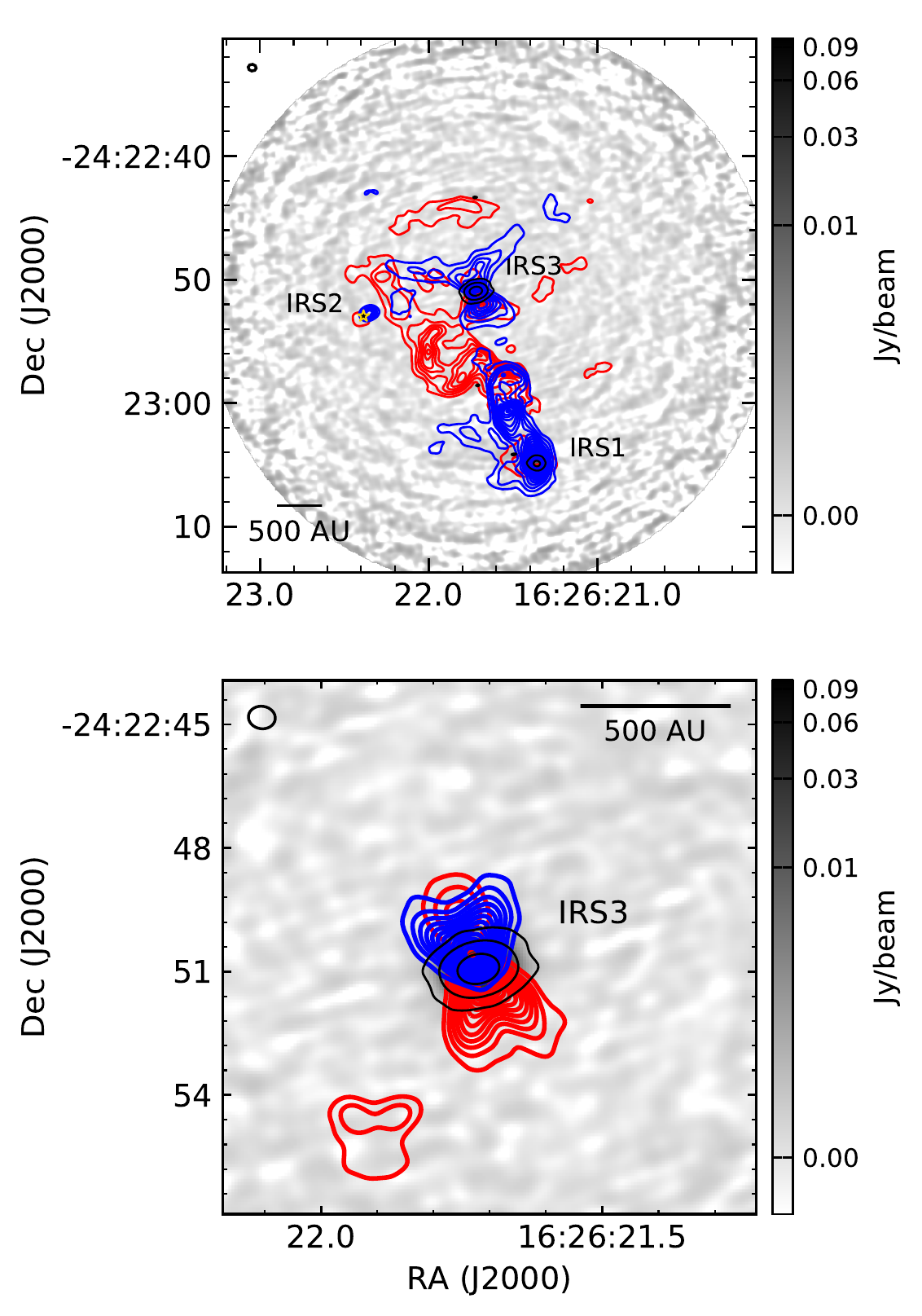}
\caption{Top: 1.1~mm continuum toward GSS 30 as in Figure \ref{fig:cont_notaper} (grayscale and black contours) with (top) red contours showing CO emission integrated from 5.5~\kms\ to 11.75~\kms\ and blue contours showing CO emission integrated from $-5.25$~\kms\ to 1.25~\kms. \revise{The star shows the position of GSS 30 IRS2.} Bottom: zoom into a smaller region around GSS 30 IRS3 showing the bipolar outflow high velocity component. The CO emission is integrated from $\pm 8 - 20$~\kms\ from the typical cloud $v_\mathrm{LSR}$ of 2.5~\kms. \label{fig:16263_2422_CO}}
\end{center}
\end{figure}

\begin{figure*}
\begin{center}
\includegraphics[height=0.45\textheight]{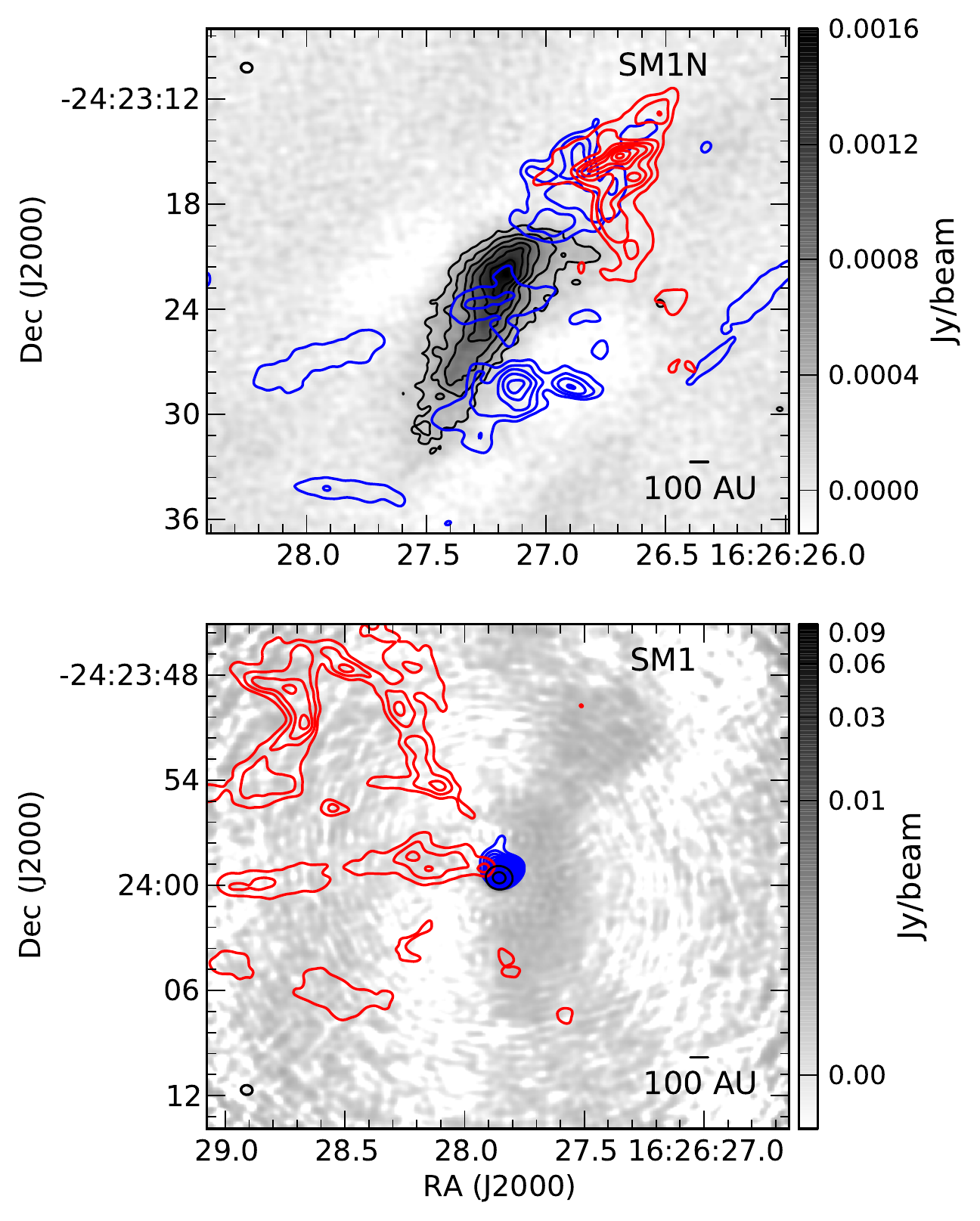}
\hfil
\includegraphics[height=0.44\textheight,trim=0 0 0 5,clip]{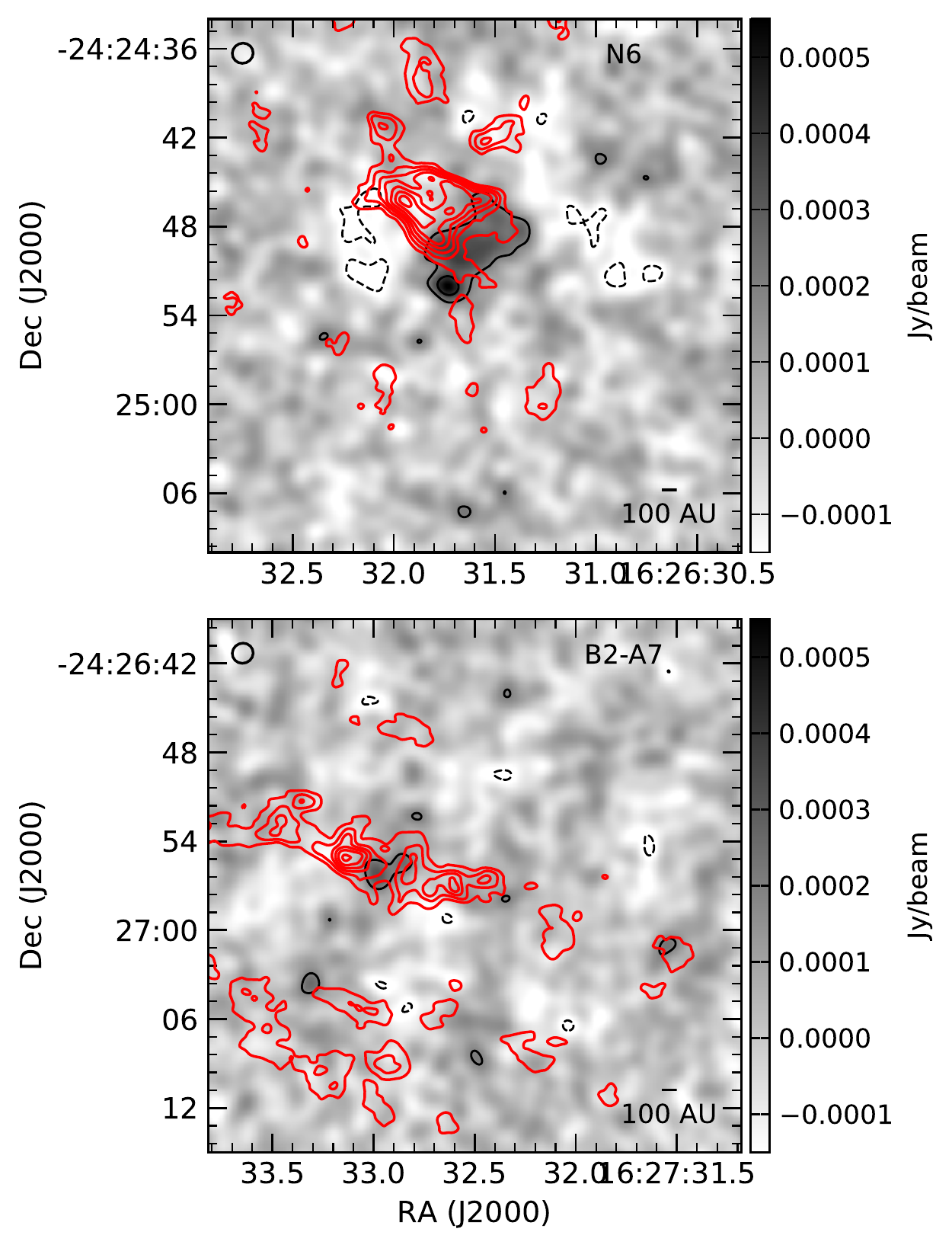}
\caption{Blue- and red-shifted CO emission (blue and red contours) toward four of the five targets with detected continuum emission. In all panels, grayscale and gray contours show 1.1~mm continuum as in Figures \ref{fig:cont_notaper} and \ref{fig:cont_taper}. 
Left top: SM1N, where red contours show CO emission integrated from 4.75~\kms\ to 9~\kms\ and blue contours show CO emission integrated from $-5$~\kms\ to 2~\kms. 
Left bottom: SM1, where red contours show CO emission integrated from 5~\kms\ to 9~\kms\ and blue contours show CO emission integrated from $-1.5$~\kms\ to 2.25~\kms. Note that we do not show N3-mm in this plot. While CO emission is seen toward the continuum source, the sensitivity is low due to its location at the edge of the ALMA primary beam. 
Right top: Oph A N6, where red contours show CO emission integrated from 5.5~\kms\ to 7.0~\kms. Toward N6, the blue-shifted CO emission is completely dominated by the VLA1623 outflow and suffers from the lack of short-spacing data, and is not shown here. 
Right bottom: B2-A7, where red contours show CO emission integrated from 6.25~\kms\ to 9.0~\kms. Toward B2-A7, very little blue-shifted CO emission is seen.
\label{fig:four_src_CO}}
\end{center}
\end{figure*}

\begin{figure*}
\begin{center}
\includegraphics[height=0.6\textheight]{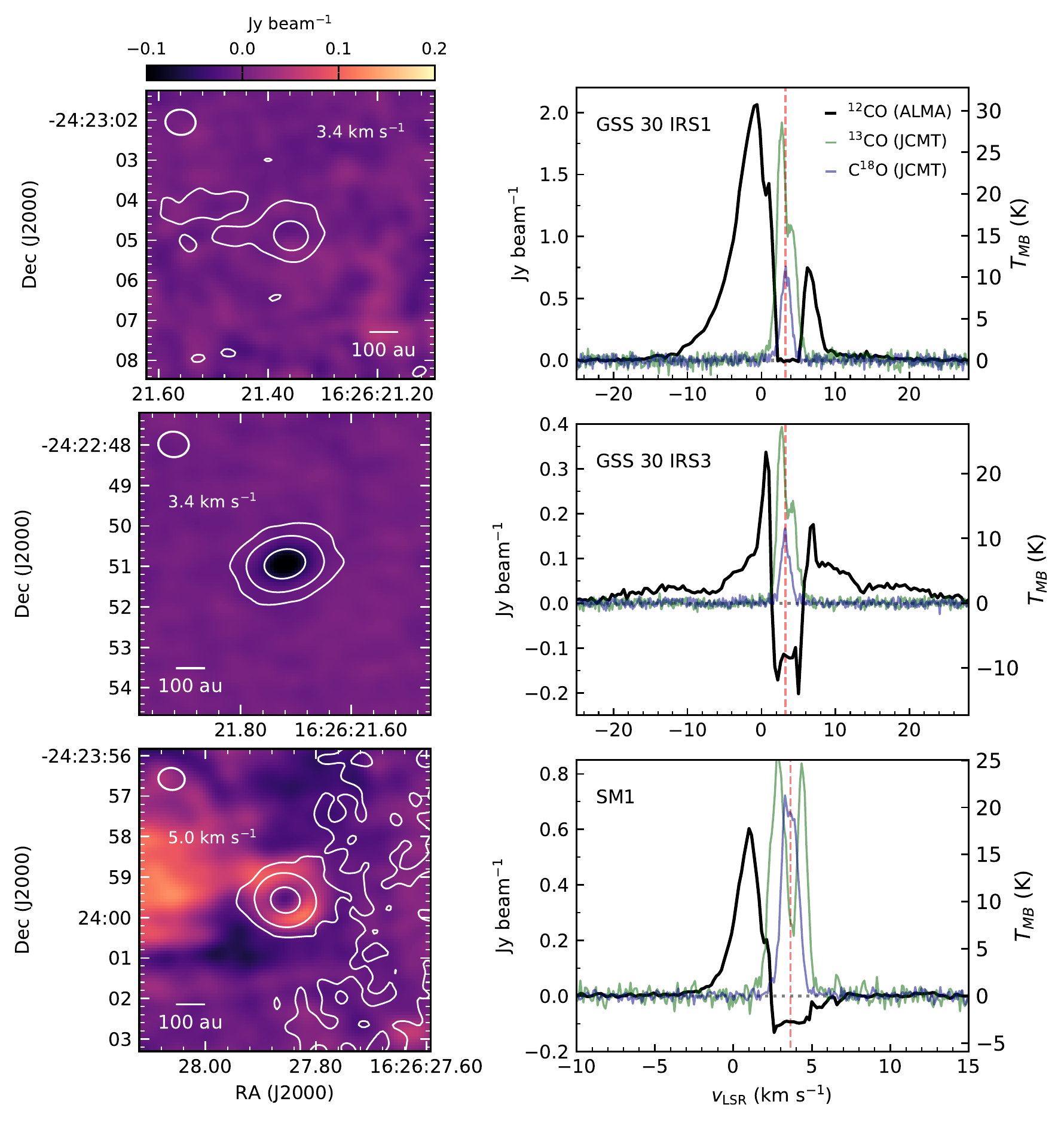}
\caption{Left: CO 2-1 emission at the indicated velocity toward GSS 30 IRS3 (top) and SM1 (bottom) where both sources are heavily self-absorbed. The colour scale is the same for both panels. White contours show 1.1~mm continuum emission as in Figure \ref{fig:cont_notaper}. The CO beam is shown at upper left. Right: ALMA CO~2-1 spectra toward the emission peak of the compact 1.1~mm continuum structure for each source (black lines). The vertical red line shows $v_\mathrm{LSR}$ of the larger-scale dense gas traced by \amm~(1,1). Broad line wings in the ALMA emission are seen toward GSS 30 IRS3, but not SM1. Light green and blue lines show the spectra of $^{13}$CO and C$^{18}$O 3-2 emission, respectively, observed at 14\arcsec\ resolution with the JCMT \citep[$T_\mathrm{MB}$ scale at right;][]{white_2015}. \label{fig:co_spectra}}
\end{center}
\end{figure*}

\subsection{CO emission}
\label{sec:co}

\subsubsection{Outflows}

CO emission is detected toward all six targets, but in most cases it suffers from severe self-absorption at the cloud \vlsr, as well as from the lack of short-spacing data. Here, we focus primarily on the CO emission blue- and red-shifted from the cloud \vlsr\ toward the sources with continuum detections, as well as on the extended CO line wings visible toward GSS 30 IRS3. 

Figures \ref{fig:16263_2422_CO} and \ref{fig:four_src_CO} show the blue- and red-shifted emission toward the five sources with continuum detections. Toward both IRS1 and IRS3 in GSS 30, clear bipolar outflows are detected. The outflow from GSS 30 IRS3 is very wide at small velocity offsets from the cloud \vlsr\ (Figure \ref{fig:16263_2422_CO}, top), but at higher velocity offsets ($\pm 8-20$~\kms) shows a much more compact and narrower component (Figure \ref{fig:16263_2422_CO}, bottom). In addition, at lower velocities, the blue-shifted emission (and to a lesser extent, the red-shifted emission) is visible on both sides of the continuum peak, but at high velocities the lobes are more clearly separated. The direction of both the wide- and narrow-angle outflow components are perpendicular to the elongation of the continuum emission. \revise{We also detect blue- and red-shifted emission at low velocities toward GSS 30 IRS2. The total extent of the emission is $\lesssim 2$\arcsec, or $\sim 275$~au. Self-absorption at the cloud \vlsr\ precludes any further analysis of this feature, although we note that the size of the CO feature is similar to that found for faint gas disks around T Tauri stars \citep{pietu_2014}.} 

SM1 (Figure \ref{fig:four_src_CO}) also shows significant blue- and red-shifted emission. While the red-shifted emission appears to illuminate a wide outflow lobe, the blue-shifted emission is strong but compact in extent, centered just offset from the continuum source in the north-west direction. 

Toward SM1N, we detect both blue- and red-shifted emission, but Figure \ref{fig:four_src_CO} shows no clear bipolar outflow, and no emission at velocities greater than a few \kms\ from the cloud velocity.

Toward N6-mm and B2-A7, we show in Figure \ref{fig:four_src_CO} only the red-shifted CO emission. For N6, the blue-shifted emission is dominated by the powerful and narrow outflow driven by VLA 1623 \citep{andre_1990,white_2015}. Red-shifted CO emission is detected extended north of the continuum peak, and may be the red lobe of a low-velocity ($\lesssim 4$~\kms) outflow. A blue lobe along this axis would unfortunately be masked by the large-scale VLA 1623 outflow. The gas velocities and the extent of the CO emission are both in agreement with some models of outflows driven by FHSCs. 

Toward B2-A7, all emission blueward of the cloud \vlsr\ is either resolved out in the 12m Array data, or self-absorbed. Red-shifted emission is detected on either side of the continuum emission detected in the tapered images, suggesting that the continuum feature is real despite its relatively low SNR.  

\subsubsection{Absorption}

We additionally find strong, compact CO absorption features toward the compact continuum components in GSS 30, IRS3 and IRS1, and SM1, where the absorption coincides precisely with the compact continuum emission. We show this absorption in Figure \ref{fig:co_spectra}, where the left panels present the CO emission at the indicated velocity, while the right panels present the CO spectrum toward each source at the continuum emission peak. 

The Figure additionally identifies the $v_\mathrm{LSR}$ of the larger-scale dense gas as traced by \amm\ (1,1) (dashed red line). We further show single-dish $^{13}$CO and C$^{18}$O $J=3-2$ spectra at the same location \citep[14\arcsec\ resolution;][]{white_2015}, where the emission peak of the more optically-thin C$^{18}$O line coincides with the \amm~(1,1) $v_\mathrm{LSR}$. 

While all sources show substantial self-absorption of the CO emission at the cloud velocity, we only detect absorption against the compact continuum toward GSS 30 IRS3 and SM1. The absorption feature is broad in velocity, and for both sources is centered near to the core \vlsr\ as measured on larger scales through \amm: \vlsr$=3.65$~\kms\ toward SM1, and \vlsr$=3.23$~\kms\ toward GSS 30 IRS3 \citep{GAS_2017}. 

Toward SM1, the CO spectrum exhibits a clear inverse P-Cygni profile suggestive of infall motions, where the red-shifted emission is seen in absorption against the bright continuum emission of the compact source.  Similar P-Cygni profiles were identified in the line emission from several dense-gas tracing molecular species toward the Class 0 protostar B335 \citep{evans_2015}, where the profiles reveal unambiguously the accretion of material onto the central luminosity source. Since CO is excited in lower density material, however, it is less clear whether the clear infall signature of the inverse P-Cygni profile is an indication of large-scale infall of the cores themselves, or accretion directly onto the compact sources. 

\revise{Following \citet{difrancesco_2001,pineda_2012}, we can fit the inverse P Cygni profile toward SM1 and find an infall velocity $v_{in} \sim 0.4$~\kms. If we assume that the CO emission profile toward SM1 is indeed tracing infall motions of the compact, ALMA-detected core, we estimate a mass accretion rate, $\dot{M} = 4~\pi~r^2 \rho~v_{in} \sim 3 \times 10^{-5}$~\msun~yr$^{-1}$, where we calculate $\rho$ and $r$ from the Gaussian fit results listed in Table \ref{tab:gaussfit}. If the infall motions toward SM1 instead originate in less dense, more extended gas, the derived mass accretion rate is an upper limit. Nevertheless, this estimate is only slightly greater than mass accretion rates calculated for B335 \citep{evans_2015} and the FHSC candidate L1451-mm \citep{maureira_2017}. Compared with Class 0 protostars, the SM1 estimate agrees well with results for the Class 0 protostar IRAS16293 in Ophiuchus \citep{pineda_2012}, but is less than that calculated for the NGC 1333 IRAS 4A and 4B \citep{difrancesco_2001}. A more accurate calculation and better comparison requires the sources to be observed in the same high density transitions, however, and modeled in a uniform manner. }

Line profiles indicative of infall are seen clearly toward GSS 30 in the JCMT spectra that trace the larger core scale. Here, the optically-thick $^{13}$CO is self-absorbed with a stronger blue peak, while the optically thin C$^{18}$O peaks at the self-absorption dip (and additionally matches the \amm\ \vlsr). Self-absorption on core scales is also seen in the $^{13}$CO spectrum toward SM1, but the line is more symmetric, and the C$^{18}$O may be optically thick as well, making an infall determination difficult. Regardless, it is clear that Figure \ref{fig:co_spectra} shows absorption strongly associated with the GSS30 IRS3 and SM1 continuum sources.

The well-developed, high-velocity bipolar outflow seen toward GSS 30 IRS3 in Figure \ref{fig:16263_2422_CO} (bottom) is clearly visible in the line wings. Meanwhile, though SM1 appears to have an extended red outflow lobe in Figure \ref{fig:four_src_CO}, we do not see any red-shifted emission in the spectrum toward the continuum peak. 

\section{Discussion}
\label{sec:disc}

\subsection{The nature of the compact sources}

We have analysed the emission from multiple compact, and in several cases elongated, continuum sources detected with ALMA toward several dense molecular cores in the Ophiuchus molecular cloud. We propose that some of these compact sources are young disks or pseudodisks around extremely young protostars. \revise{In Table \ref{tab:summary}, we list the properties of the ALMA targets and their likely identification as starless cores, first hydrostatic cores, or protostellar. Below, we briefly summarize the results for each source.} 

Toward the previously known Class I source GSS 30 IRS3, we find a likely compact disk $\sim 70$~au in size and a well-established, high-velocity bipolar outflow in CO, revealing an actively accreting protostar. The clear accretion activity, however, belies the interpretation of the previously measured, low $L_\mathrm{bol}$ value as the result of low accretion activity. The high extinction toward GSS 30 IRS3 likely complicates $L_\mathrm{bol}$ measurements. GSS 30 IRS1 also shows classic disk and outflow structure in the ALMA data. \revise{The spectral indices of the continuum emission toward both sources indicate either grain growth, or high optical depth, either of which support their identification as disk structures.}

\revise{We also detect compact continuum and extended CO emission toward SM1. Like the sources toward GSS 30, the continuum emission appears either optically thick, or is indicative of grain growth. The compact continuum source toward} SM1 is not elongated, however, suggesting that if the continuum emission originates in an accretion disk, we are viewing it nearly face-on. The variable hard x-ray detection toward SM1 reveals the presence of an actively accreting protostar. In contrast \revise{with GSS 30 IRS3}, the outflow detected toward SM1 is lower velocity and more compact, indicating that SM1 is less evolved.  In addition, the inverse P-Cygni line profile of the CO emission toward SM1 strongly suggests infall, either of the larger-scale core, or closer to the disk itself. \revise{We estimate a mass accretion rate $\dot{M} \sim 3 \times 10^{-5}$~\msun~yr$^{-1}$ if the infall motions originate close to the continuum source. Faint extended continuum emission is also seen toward SM1, although it appears offset from the compact source and aligned with the larger-scale Oph A ridge.}

\revise{The newly-detected compact continuum source toward N3-mm is another disk-like source, with either high optical depth in the millimeter-submillimeter continuum, or significant grain growth. Similarly to SM1, the variable hard x-ray detection toward N3-mm indicates the presence of an embedded, accreting protostar.}
We do not detect an outflow toward N3-mm, but this may be due to a lack of sensitivity at the edge of the ALMA primary beam if the outflow is compact and low-velocity. Given the position and orientation of the N3-mm continuum feature, one lobe would likely extend into the primary beam of the ALMA observations if it were powering a larger-scale outflow \revise{similar to GSS 30 IRS3 or IRS1}. Both SM1 and N3-mm thus appear to be younger protostars than GSS 30 IRS3 and IRS1. 

\revise{We detect here for the first time a compact continuum source of size $\sim 100$~au toward Oph A N6.} In contrast to the continuum sources detected toward SM1 and GSS 30, \revise{the N6} compact source is embedded within a larger, extended continuum structure a few hundred au in size, with mass $M \sim 0.025 - 0.03$~\msun. \revise{The extended continuum emission seen with ALMA agrees with previous millimeter continuum measurements at lower resolution \citep{bourke12}. To the north of the N6 compact source, we detect CO emission red-shifted by only a few \kms\ in velocity from the cloud \vlsr. Given the position angle of the emission relative to the continuum source, however, any blue-shifted counterpart would be confused with the strong, narrow outflow from VLA 1623.} 

\revise{Toward SM1N, we detect at 1.1~mm an elongated, extended continuum structure first identified in \papi, with total mass $M \sim 0.03$~\msun\ and mean density $\sim 10^7$~\cc\ assuming typical dust properties for dense cores. Analysis of the uv-amplitudes shows complex structure, however, and suggests the presence of an unresolved emission source. Fitting the radial emission profile of the continuum image, we find a Gaussian width of 1.44\arcsec\ $\pm 0.02$\arcsec, or $197 \pm 3$~au at the distance to Ophiuchus. Assuming the core is isothermal, the density profile decreases as $r^{-1.3}$ at $r \lesssim 200$~au, turning over to $r^{-2}$ at larger radii.} Red- and blue-shifted CO emission is detected toward SM1N, but extends only a few \kms\ in velocity from the cloud \vlsr\ and does not show clear bipolar features. Any outflows are likely low-velocity, similar to predictions from FHSC models, or highly inclined along the plane of the sky.

In addition, both SM1N and N6-mm show offsets between \revise{the peak locations of} continuum \revise{emission} and \revise{emission from} deuterated species \citep[\papi;][]{pon_2009,bourke12}. The detection of the compact source N6-mm, if an extremely young protostar, might explain the observed offset between the continuum peak and the location of the maximum value of the deuterium fractionation in the core. \citeauthor{bourke12} show that the maximum \ddia\ to \dia\ ratio lies to the northwest of both the continuum and \dia\ emission peaks by $\sim 5$\arcsec\ ($\lesssim 700$~au). Heating by a deeply embedded protostar can decrease the nearby deuterium fractionation both by liberating CO off nearby dust grains, which then react with deuterated molecules and their parent species, and by increasing the local temperature sufficiently that the deuterium-enhancing chemical reactions are no longer favoured.

At lower resolution than presented here, the extended emission toward SM1N and N6-mm is characterized by small non-thermal velocity dispersions \citep[$\sigma_{\mathrm{NT}} \sim 0.08$~\kms;][]{bourke12,friesen14}. Regardless, given the small masses of the continuum structures, the ratio of the observed mass to the virial mass \citep[following ][]{bertoldi_1992} suggests that they may not be formally unstable by a virial criterion. While the narrow line widths observed suggest any infall or rotation motions must be small, sensitive high resolution line observations of dense gas tracers are needed to identify infall and/or rotation to better evaluate the stability of these objects. 

\citet{commercon_2012} simulate ALMA observations of model FHSCs and pseudodisks. The predicted peak flux density of their models, when scaled to match the wavelength of our observations given typical dust properties, is generally greater by a factor of a few than we find toward SM1N, N3-mm, or N6. The peak flux densities of SM1 and GSS 30 IRS3 are a better match to the models, but their well-established bipolar outflows suggest that the responsible embedded sources are actively-accreting protostars that are past the FHSC stage. Note also that GSS 30 IRS3 exhibits CO line wings that trace outflow velocities $\gtrsim 20$\ \kms, much greater than predicted for outflows driven by FHSCs. 

\subsection{\revise{The actively star-forming Ophiuchus A core}}

Single-dish (sub)millimeter observations revealed that the Ophiuchus A ridge contains a string of compact, dense cores \citep{andre93}, in addition to the prototypical Class 0 protostar VLA 1623 \citep{andre_1990}. High resolution observations of dense gas tracers, such as \dia, further identified cores like N6 that are characterized by high density and low velocity dispersion \citep{difrancesco_2004,pon_2009,bourke12}. ALMA observations presented here and in previous work \citep{friesen14,kirk_2017} have revealed multiple compact protostellar sources in Oph A that were previously unseen in far-infrared and short-wavelength submillimeter observations. Furthermore, we have detected additional cores within Oph A that are likely on the cusp of collapse. Thus far, of the dense cores identified within Oph A through the various observations listed above, only SM2 \citep[located between SM1 and N6;][]{andre93,difrancesco_2004} has no evidence of an embedded protostar or FHSC. Including the two Class 0 components of the triple system associated with VLA 1623 \citep{murillo_2013b}, and assuming all the embedded sources are Class 0 and younger, the $\sim 30$\ \msun\ Oph A ridge is in the process of forming at least six (VLA 1623 A and B, SM1, SM1N, N3-mm, N6) low-mass stars within the expected Class 0 lifetime of $\sim 0.1 - 0.16$~Myr \citep{evans_2009}. 

\floattable 
\begin{deluxetable}{lccccccc} 
\tabletypesize{\footnotesize} 
\tablecolumns{8} 
\tablewidth{0pt} 
\tablecaption{\revise{Summary of source properties} \label{tab:summary}} 
\tablehead{ 
\colhead{Source} & \colhead{Compact continuum} & \colhead{M$_\mathrm{comp}$} & {Extended continuum} & \colhead{Outflow} & \colhead{Infall}  & \colhead{Disk} & \colhead{Evolutionary} \\
\colhead{} & \colhead{detection\tablenotemark{a}} & \colhead{(\msun)} & \colhead{emission\tablenotemark{a}} & \colhead{} & \colhead{} & \colhead{} & \colhead{stage\tablenotemark{b}}
} 
\startdata 
GSS 30 IRS1 &  Y & 0.0005 - 0.0043 & N & Y & ? & Y & P \\ 
GSS 30 IRS3 &  Y & 0.0046 - 0.0446 & N & Y & ? & Y & P \\ 
SM1N & Y & 0.03 & Y & ? & N & N & SC/FHSC? \\
N3-mm & Y & 0.0021 - 0.0130 & N & N\tablenotemark{c} & N & Y & P \\ 
SM1 &  Y & 0.0042 - 0.0404 & Y & Y & Y & Y & P \\ 
N6-mm & Y & 0.0002 & Y & ? & N & ? & FHSC?/P \\ 
16267-2417 & N & \nodata & N & N & N & N & SC \\
B2-A7 & ? & \nodata & N & N & N & N & SC \\
\enddata 
\tablenotetext{a}{Defined as continuum emission detected by ALMA in the observations presented here.}
\tablenotetext{b}{Evolutionary stages: SC - starless core; FHSC - first hydrostatic core; P - protostellar.}
\tablenotetext{c}{Source is near the edge of the primary beam; a large outflow is ruled out but a compact outflow may still be present.}
\end{deluxetable}

\section{Summary and Outlook}
\label{sec:summ}

In summary, we have observed with ALMA the millimeter continuum and CO 2-1 emission toward six cores in L1688. We find the following:

\begin{itemize}
\item GSS 30: Two compact, sub-arcsecond continuum sources detected (IRS1 and IRS3), with spectral indices that indicate either dust grain growth or that the continuum emission is optically thick. Both show well-developed, bi-polar outflows in CO~2-1 emission, while GSS 30 IRS3 also has a more narrow, high-velocity component \revise{that extends to $\pm 20$~\kms\ from the core \vlsr. The T Tauri protostar GSS 30 IRS2 is undetected in continuum, but detected in CO emission.}
\item SM1: Two compact, sub-arcsecond continuum sources detected (SM1, N3-mm), also with spectral indices that indicate either dust grain growth or that the continuum emission is optically thick. A \revise{low velocity ($\lesssim 6$~\kms),} bi-polar, but asymmetric, outflow is seen toward SM1 in CO~2-1 emission. \revise{The CO emission toward the SM1 continuum peak shows an inverse P-Cygni profile indicating infall, and we calculate a mass accretion rate $\dot{M} \sim 3 \times 10^{-5}$~\msun~yr$^{-5}$ assuming the infall arises close to the compact continuum source.} No outflow is detected toward N3-mm, but this may be limited by the ALMA primary beam. 
\item SM1N: \revise{We find a highly centrally concentrated,} extended continuum \revise{structure with mass $M \sim 0.03$~\msun\ and density profile that follows a power law of $r^{-1.3}$. Evidence for additional compact structure is seen in the uv-data}. The observed spectral index is consistent with typical measurements in the dense ISM. Both blue- and red-shifted CO emission features are present near the continuum emission, but do not clearly delineate an outflow. 
\item N6: A sub-arcsecond continuum source is detected \revise{(N6-mm)}, along with extended continuum emission. Red-shifted CO may indicate the presence of a compact outflow, but we are unable to detect any blue-shifted CO emission due to the impact of the VLA 1623 outflow. 
\item 16267-2417: While high-resolution \amm\ observations suggest the presence of a peaked continuum source, no continuum or CO emission was detected with ALMA to our sensitivity limits.
\item B2-A7: Faint, compact continuum emission detected at the $\sim 4\times\sigma$ level. Red-shifted CO emission is present near the continuum peak, but no clear outflow is identified. 
\end{itemize}

SM1N and N6-mm are candidates to be the youngest protostars or first hydrostatic cores in the Ophiuchus molecular cloud. The continuum sources identified toward GSS 30 and SM1 are likely compact disks around young protostars. Future high resolution observations, like those available with ALMA, will better characterize these intriguing objects. In particular, resolved continuum observations over multiple wavelengths will allow detailed modeling of the physical and temperature structure of the compact sources and disks, including the effects of the dust opacity, providing more precise measurements of the source masses. High-resolution line observations of dense gas tracers will discern Keplerian motions and infall, while observations of outflow tracers that include short-spacing information would enable robust measurements of the outflow sizes and momenta, to better compare with theoretical models of the evolution of FHSCs and young protostars. 

\acknowledgments

We thank the referee for their helpful suggestions that improved this manuscript. Partial salary support for A. P. was provided by a Canadian Institute for Theoretical Astrophysics (CITA) National Fellowship. PC and JEP acknowledge support from the European Research Council (ERC; project PALs~320620). JKJ also acknowledges support from the European Research Council through the ERC Consolidator grant ``S4F" (grant agreement No~646908). This paper makes use of the following ALMA data: ADS/JAO.ALMA{\#} 2013.1.00937.S. ALMA is a partnership of ESO (representing its member states), NSF (USA), and NINS (Japan), together with NRC (Canada) and NSC and ASIAA (Taiwan), in cooperation with the Republic of Chile. The Joint ALMA Observatory is operated by ESO, AUI/NRAO, and NAOJ. The Australia Telescope Compact Array is part of the Australia Telescope National Facility which is funded by the Australian Government for operation as a National Facility managed by CSIRO. The National Radio Astronomy Observatory is a facility of the National Science Foundation operated under cooperative agreement by Associated Universities, Inc. 

\facilities{ALMA, ATCA}
\revise{
\software{CASA \citep{mcmullin_2007}, MIRIAD \citep{sault95}, Astropy \citep{astropy_2018}, APLpy \citep{robitaille_2012}}}

\bibliographystyle{apj}
\bibliography{biblio}

\end{document}

%% file: mydefs.tex
\newcommand{\amm}{NH$_3$}
\newcommand{\kms}{km\,s$^{-1}$}
\newcommand{\cc}{cm$^{-3}$}
\newcommand{\vlsr}{$v_{\mathrm{LSR}}$}

\newcommand{\tkin}{$T_\mathrm{K}$}

\newcommand{\hd}{H$_2$D$^+$}
\newcommand{\dia}{N$_2$H$^+$}
\newcommand{\ddia}{N$_2$D$^+$}

\newcommand{\lint}{$L_\mathrm{int}$}

\newcommand{\msun}{M$_\odot$}
\newcommand{\papi}{Paper I}
\newcommand{\muJy}{$\mu$Jy}